\documentclass[usenatbib]{mn2e}  
\pdfoutput=1 
\usepackage[pdftex]{graphicx} 
\usepackage{times}
\usepackage{rotate} 
\usepackage{amssymb} 
\usepackage{rotating}
\usepackage{epstopdf} 
\usepackage{graphicx}
\usepackage{array}

\title[Dark   matter   halo   mass  measurements   using   photometric
redshifts]{Large   scale  clustering  measurements   with  photometric
redshifts:  comparing   the  dark  matter  halos  of   X-ray  AGN, 
star-forming and passive galaxies at $z\approx1$}

\author[Georgakakis et al.]{A. Georgakakis$^{1,2}$, G. Mountrichas$^2$, M. Salvato$^1$, D. Rosario$^1$,  P. G. P\'erez-Gonz\'alez$^{3, 4}$, 
\newauthor D. Lutz$^1$, 
K. Nandra$^1$, A. Coil$^{5}$, M. C. Cooper$^{6}$, J. A. Newman$^{7}$, S. Berta$^1$, B. Magnelli$^1$, 
\newauthor  P. Popesso$^1$,  F. Pozzi$^8$
\\\\
$^1$Max Planck  Institut f\"{u}r Extraterrestrische  Physik, Giessenbachstra\ss e, 85748 Garching, Germany\\ 
$^2$National Observatory of Athens, V.  Paulou  \& I.  Metaxa, 11532,  Greece\\ 
$^3$Departamento de Astrof\'{\i}sica, Facultad de CC. F\'{\i}sicas, Universidad Complutense de Madrid, E-28040 Madrid, Spain\\
$^4$Steward Observatory, The University of Arizona, 933 N. Cherry Ave., Tucson, AZ 85721, USA\\
$^5$Center for Astrophysics and Space Sciences, University of California, San Diego, 9500 Gilman Dr., La Jolla, CA 92093-0424, USA\\
$^{6}$Center for Galaxy Evolution, Department of Physics and Astronomy,
University of California, Irvine, 4129 Frederick Reines Hall Irvine, CA 92697 USA\\
$^{7}$Department of Physics and Astronomy \& Pittsburgh Particle
Physics, Astrophysics and Cosmology Center (PITT PACC), University of Pittsburgh, \\Pittsburgh, PA 15260, USA\\
$^8$Dipartimento di Fisica e Astronomia, Universita di Bologna, via Ranzani 1, I-40127 Bologna, Italy
}

\begin{document}

\maketitle

\label{firstpage}

\begin{abstract} We  combine multiwavelength data in  the AEGIS-XD and
C-COSMOS surveys  to measure  the typical dark  matter halo mass  of X-ray
selected AGN [$L_X( \rm 2 - 10 \, keV )> 10^{42} \, erg \, s^{-1}$] in
comparison with  far-infrared selected star-forming  galaxies detected
in the Herschel/PEP survey (PACS Evolutionary Probe; $L_{IR}>10^{11}\,
L_{\odot}$) and quiescent systems  at $z\approx1$.  We develop a novel
method  to measure  the clustering  of extragalactic  populations that
uses  photometric  redshift   Probability  Distribution  Functions  in
addition  to  any spectroscopy.   This  is  advantageous  in that  all
sources in  the sample are used  in the clustering  analysis, not just
the subset with secure spectroscopy.   The method works best for large
samples.  The loss of accuracy  because of the lack of spectroscopy is
balanced  by increasing  the number  of  sources used  to measure  the
clustering.    We   find  that   X-ray   AGN,  far-infrared   selected
star-forming  galaxies and  passive systems  in the  redshift interval
$0.6<z<1.4$   are   found   in    halos   of   similar   mass,   $\log
M_{DMH}/(M_{\odot}\,h^{-1})\approx13.0$.    We  argue  that   this  is
because the galaxies in all three samples (AGN, star-forming, passive)
have similar stellar mass  distributions, approximated by the $J$-band
luminosity.  Therefore  all galaxies  that can potentially  host X-ray
AGN, because they  have stellar masses in the  appropriate range, live
in dark matter haloes of $\log M_{DMH}/(M_{\odot}\,h^{-1})\approx13.0$
independent  of their  star-formation rates.   This suggests  that the
stellar mass  of X-ray  AGN hosts is  driving the  observed clustering
properties of this population.   We also speculate that trends between
AGN  properties (e.g.   luminosity,  level of  obscuration) and  large
scale environment may be related to differences in the stellar mass of
the host galaxies.
\end{abstract}


\section{Introduction}

In recent  years progress  has been made  in our understanding  of the
physical conditions  under which  supermassive black holes  (SMBHs) at
the centres of galaxies grow their masses.  Large extragalactic survey
programs   combining   information  from   different   parts  of   the
electromagnetic spectrum made possible  the study of the properties of
the galaxies  that host Active  Galactic Nuclei (AGN),  which signpost
accretion events onto SMBHs.  As a result constraints have been placed
on   e.g.   the   morphology   \citep{Georgakakis2009,  Cisternas2011,
Kocevski2012},    stellar    mass    distribution    \citep{Bundy2008,
Georgakakis2011,   Aird2012}   and   position   on  the   cosmic   web
\citep[e.g.][]{Coil2009, Mountrichas2012, Mountrichas2013, Krumpe2012,
Allevato2012, Krumpe2013_review, Komiya2013} of  the hosts of AGN over
a range of redshifts and accretion luminosities.  These diagnostics of
the physical conditions  on large scales (kpc and  Mpc) have also been
related  to  the accretion  properties  of  the  SMBH, e.g.   specific
accretion  rate   or  Eddington  ratio  \citep[e.g.][]{Schawinski2010,
Aird2012, Aird2013},  to better understand  what triggers AGN  and how
they affect their immediate environment.

The general picture emerging from these  studies is that at least in a
statistical sense,  star-formation episodes are related  to the growth
of  SMBHs.  The  star-formation  rate of  galaxies  for example,  when
integrated  over cosmological  volumes, evolves  with redshift  in the
same manner as the  AGN accretion density \citep{Zheng2009, Aird2010}.
Similarly,  the mean  specific  star-formation rate  of galaxies  also
appears to  follow the  same evolution pattern  as the  AGN population
\citep{Georgakakis2011,  Santini2012,   Mullaney2012}.   Stacking  the
far-infrared fluxes at  the positions of AGN shows  that these systems
at any  given redshift lie, on  the average, on or  perhaps even above
the  main  star-formation  sequence  of  galaxies  \citep{Santini2012,
Mullaney2012, Rovilos2012, Rosario2013}. At  the same time however, it
has been  become clear that in  individual AGN there  is no one-to-one
correspondence between the level  of star-formation in the host galaxy
and   the    accretion   luminosity   \citep{Shao2010,   Mullaney2012,
Rosario2012}.    There   is   rather   substantial  scatter   in   the
star-formation/accretion-luminosity  diagram  of  AGN.   This  can  be
interpreted as a manifestation of the long term variability of AGN and
ultimately the  different timescales of star-formation  and black hole
growth \citep{Hickox2013}.

The large  scatter in the star-formation properties  of individual AGN
could  also be  understood in  the context  of different  SMBH fueling
modes that operate  in galaxies with distinct cold  gas reservoirs and
hence,  star-formation histories.   Evidence  for a  dichotomy in  the
accretion rate distribution of AGN based on the star-formation history
of   their    hosts   is    found   at   low    redshifts,   $z\la0.1$
\citep{Kauffmann_Heckman2009}.  AGN associated  with the most actively
star-forming  galaxies  have  high  Eddington  ratios  that  follow  a
log-normal  distribution.   In  contrast,  active SMBHs  in  quiescent
galaxies  are   characterised  by   low  Eddington  ratios   that  are
distributed as  a power-law.  Evidence  also exists that  these trends
between  the level  of star-formation  of galaxies  and  the Eddington
ratio of  AGN persist to higher redshift,  $z\approx1$ (Georgakakis et
al.   2014  in  prep.; but  see  Aird  et  al.  2012).   The  observed
large-scale  clustering properties  of X-ray  AGN are  also consistent
with  models  \citep{Fanidakis2012} that  postulate  two channels  for
growing SMBHs,  each one  of which takes  place in galaxies  with very
different star-formation histories \citep{Fanidakis2013}.

In  this  paper we  explore  the  relation  between AGN  activity  and
star-formation by measuring the typical dark matter halo mass of X-ray
AGN  in   comparison  with  star-forming  galaxies   detected  in  the
far-infrared by  the {\it Herschel}  space telescope.  If the  bulk of
the black  hole growth  is related to  star-formation events  then one
might expect similar  large scale environments for X-ray  AGN and {\it
Herschel}  sources.   Moreover,  if  there is  an  AGN  sub-population
associated with passive and strongly  clustered hosts we might be able
to identify its signatures in large scale clustering measurements.

We address these questions by developing a novel clustering estimation
method  that uses photometric  redshifts, in  the form  of probability
distribution functions, in addition  to any available spectroscopy, to
estimate  the  projected  correlation  function of  any  extragalactic
population.   This  is motivated  by  the  significant improvement  in
recent  years in  the  quality and  quantity  of photometric  redshift
estimates for both AGN and galaxies. The method is geared toward large
sample  sizes   to  minimise   the  impact  of   photometric  redshift
uncertainties onto the clustering  signal. It is therefore well suited
for measurements of the clustering of AGN in e.g.  the eROSITA All Sky
Survey  \citep{Merloni2012}.   One  of   the  advantages  of  the  new
clustering estimator is  that one can use in  the analysis all sources
(e.g.  X-ray  AGN, {\it Herschel} galaxies)  with optical counterparts
in  a  sample,  not  just   the  optically  brighter  ones  for  which
spectroscopy is available.  This method extends clustering measurement
techniques based on photometric redshift PDFs developed and/or applied
to  data   by  \cite{Myers2009},  \cite{Hickox2011,   Hickox2012}  and
\cite{Mountrichas2013}.    These  methods   are   geared  toward   the
determination  of  the   projected  {\it  cross}-correlation  function
between  two  extragalactic   populations  and  require  spectroscopic
redshifts  for at  least one  of the  two samples.   In  contrast, the
method presented here allows clustering measurements (auto-correlation
or cross-correlation) via the  projected correlation function, even at
the  limiting case  of no  spectroscopic  information for  any of  the
galaxy   populations  involved.   Throughout   this  paper   we  adopt
$H_0=100$\,km\,s$^{-1}$\,Mpc$^{-1}$,         $\Omega_m=0.3$        and
$\Omega_\Lambda=0.7$ and  $\sigma_8=0.8$. Rest frame  quantities (e.g.
luminosities,   dark   matter  halo   masses)   are  parametrised   by
$h=H_0/100$, unless otherwise stated.

\begin{figure}
\begin{center}
\includegraphics[height=0.9\columnwidth]{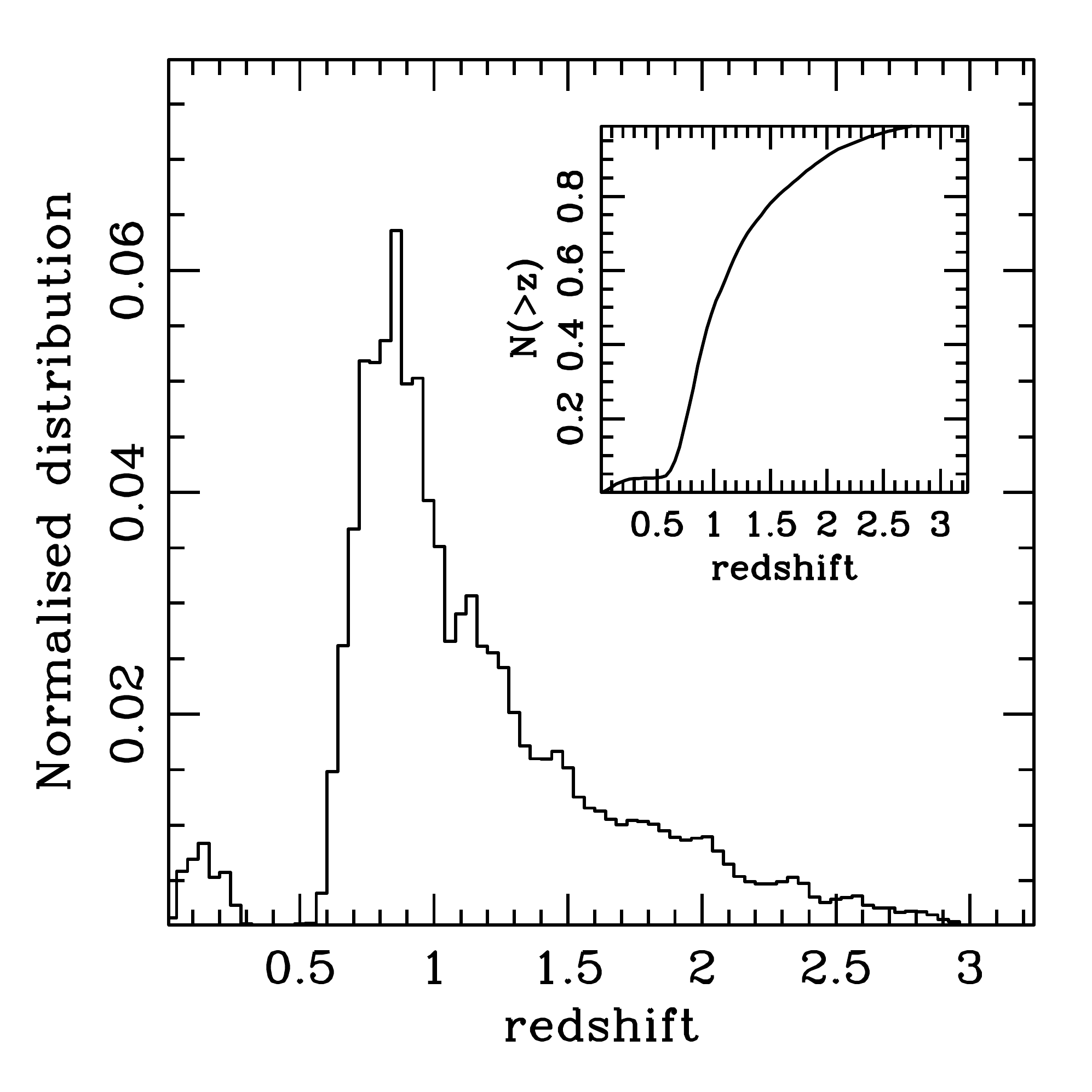}
\end{center}
\caption{Normalised redshift distribution  of galaxies in the combined
C-COSMOS and  CFHTLS-D3 fields after  applying the $B  - R$ and $R  - I$
colour cuts defined by Newman  et al. (2012) to pre-select galaxies at
$z  >  0.6$.  The  magnitude  limit  of the  two  samples  is  set  to
$R=24.5$\,mag.  The redshift distribution  is estimated by summing the
photometric redshift  PDFs of individual  galaxies.  The corresponding
cumulative redshift distribution is shown  in the inset plot. It shows
that about 80\% of the sample is in the redshift range 0.6-1.4. 
}\label{fig_nz}
\end{figure}

\begin{figure}
\begin{center}
\includegraphics[height=0.8\columnwidth]{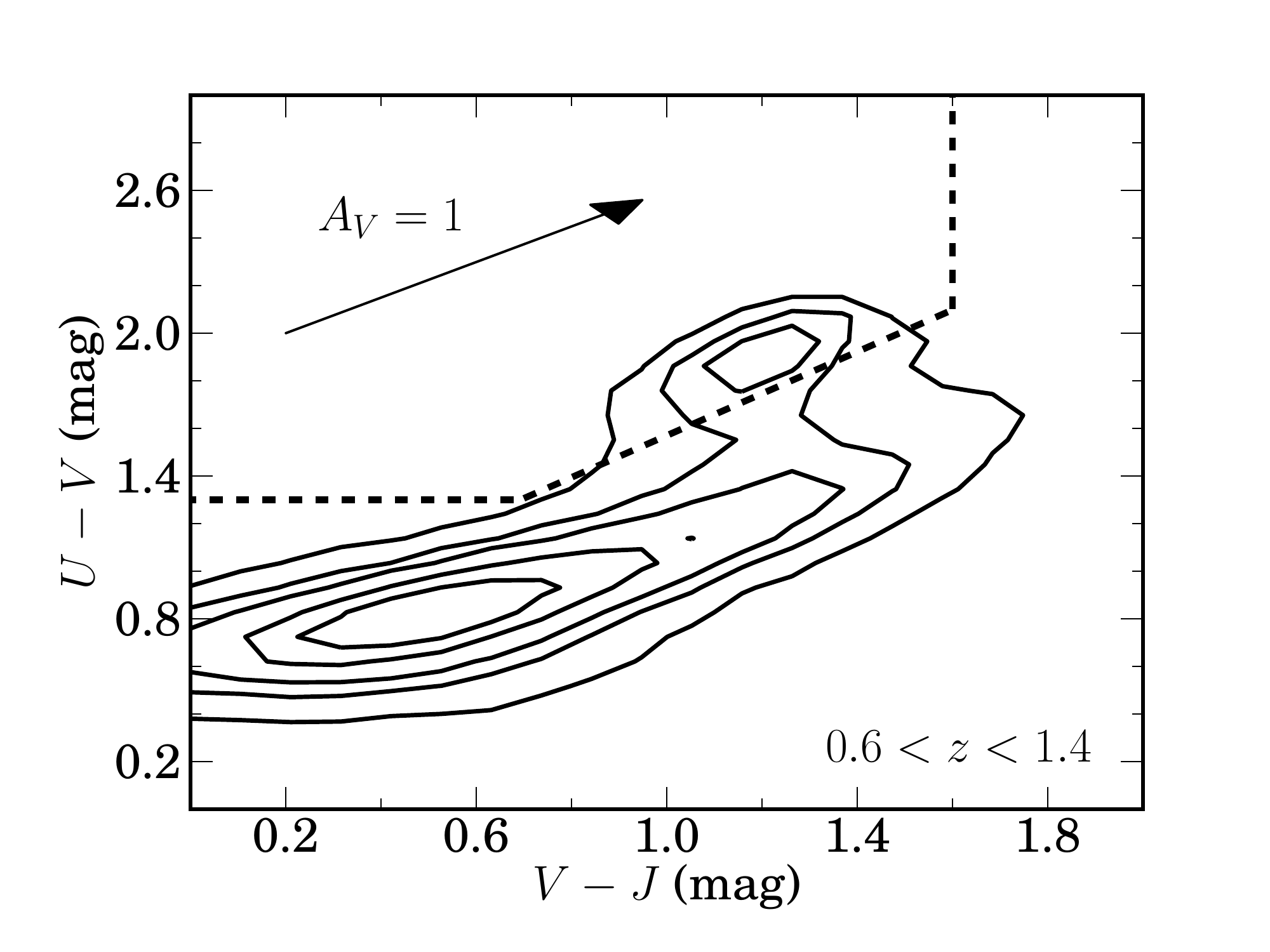}
\end{center}
\caption{$U-V$ vs  $V-J$ diagram of  galaxies (black contours)  in the
AEGIS-XD and C-COSMOS fields with $R<24.5$ and either spectroscopic or
photometric redshift estimates. For sources with photometric redshifts
we  use the  full PDFs  to  determine the  corresponding $U-V$,  $V-J$
probability  distribution  functions.   The different  contour  levels
correspond to 1000, 700, 400, 200 and 100 galaxies within bins of size
0.1\,mag.   The  dashed  lines correspond  to  $U-V=0.88\,(V-J)+0.69$,
$U-V>1.3$,   $V-J<1.6$  (Williams   et  al.    2009).    Galaxies  are
distributed  into  two   distinct  populations,  i.e.   quiescent  and
star-forming.   The  wedge, as  defined  above,  marks the  transition
region  between these  two galaxy  populations.  The  arrow  shows the
reddening vector with $A_V=1$ for the \protect\cite{Calzetti2000} law.
This  is parallel  to  the quiescent  galaxy  selection wedge.   Dusty
star-forming galaxies are  therefore separated from quiescent systems.
}\label{fig_UVJ}
\end{figure}

\section{The data} \label{section:samples}

Two extragalactic  survey fields are used to  determine the clustering
of X-ray AGN  and infrared selected galaxies at  $z\approx1$.  The All
Wavelength     Extended    Groth     strip     International    Survey
\citep[AEGIS,][]{Davis2007}  and  the  Cosmological  evolution  Survey
\citep[COSMOS,][]{Scoville2007}.  The choice of fields is motivated by
the availability  of (i)  deep Chandra and  {\it Herschel}  data, (ii)
extensive  follow-up  spectroscopic  programs  targeting  specifically
X-ray  sources and  (ii)  deep multiwavelength  imaging (UV,  optical,
infrared)  for the determination  of photometric  redshift Probability
Distribution Functions (PDFs) for  galaxies and AGN. In the subsequent
analysis  we focus  on the  part  of the  AEGIS field  which has  been
surveyed by  Chandra for a total  of 800\,ks (AEGIS-XD,  Nandra et al.
in  prep).  In  the COSMOS  field  we use  the region  covered by  the
Chandra  observations performed  between November  2006 and  June 2007
\citep[C-COSMOS,][]{Elvis2009}. Combining clustering measurements from
two extragalactic survey fields gives a better handle on the impact of
cosmic variance on the results.

\subsection{The galaxy samples} \label{section:galaxies}

The  AEGIS-XD field lies  within the  D3 region  of the  deep synoptic
Canada-France-Hawaii  Telescope Legacy  Survey (CFHTLS).   The optical
photometry ($ugriz$  bands) of the  T0004 data release is  used, which
includes photometric redshift measurement and their corresponding PDFs
with   an   estimated  dispersion   at   the   limit  $i<24$\,mag   of
$\sigma_{\Delta  z/(1+z)}\approx0.03$ \citep[][]{Coupon2009}.  Regions
of   unreliable   photometry    (CFHTLS   catalogue   parameter   {\sc
flag\_terapix$>1$})  because e.g.  of  contamination by  bright stars,
are masked  out.  In the analysis  we only use  CFHTLS optical sources
classified  as  galaxies  (CFHTLS  parameters {\sc  object}  and  {\sc
flag\_terapix}  equal  to zero),  with  reliable photometric  redshift
estimates (CFHTLS parameter {\sc zp\_reliable}$\neq-99$).

For  the C-COSMOS  field we  use the  version 1.8  of  the photometric
redshift catalogue  of \cite{Ilbert2009}, which includes  PDFs for the
photometric  redshifts and  is based  on  an improved  version of  the
photometry originally presented  by \cite{Capak2007}.  The accuracy of
the   photometric   redshifts   to  $i=24$\,mag   is   $\sigma_{\Delta
z/(1+z)}\approx0.01$ \citep{Ilbert2009}.  Spatial masks defined in the
$B$, $V$, $i$ and $z$ photometric bands have been used to identify and
exclude from the analysis sources which lie in regions with unreliable
photometry.  Optical sources with Spectral Energy Distributions (SEDs)
that  are best-fit  by stellar  templates \citep{Ilbert2009}  are also
excluded.
 
In  the  clustering  analysis  presented  in the  later  sections  the
photometric  galaxy samples in  the AEGIS-XD  and C-COSMOS  fields are
limited  to  redshifts $z\ga0.6$.   The  $B-R$  and $R-I$  photometric
criteria      for      redshift      pre-selection     defined      by
\citet[i.e. $B-R<2.35\,(R-I)-0.45$, $R-I>1.15$, $B-R<0.5$]{Newman2012}
are adopted  to exclude  galaxies below $z\approx0.6$.   The resulting
redshift distribution  is plotted in Fig.   \ref{fig_nz}.  Nearly 80\%
of the galaxy  sample lies in the redshift  interval $0.6<z<1.4$.  For
the  application of  the  above colour  cuts  the AEGIS-XD/CFHTLS  and
C-COSMOS  filtersets are  transformed to  the $BRI$  photometric bands
used   by  \cite{Newman2012}  following   the  methods   described  in
\cite{Mountrichas2013}.   Unless otherwise  stated the  galaxy samples
used in the following sections are also limited to the magnitude range
$18<R<24.5$\,mag    to    ensure    reliable   photometric    redshift
determinations  and  PDFs   \citep{Ilbert2009,  Coupon2009}.   In  the
clustering measurements described below only the part of the galaxies'
photometric redshift PDF in the range $0.6<z<1.4$ is used.

\subsection{Far-infrared galaxies} \label{section:galaxies}

{\it  Herschel}   far-infrared  (far-IR)   data  are  from   the  PACS
Evolutionary  Probe   \citep[PEP;][]{Lutz2011}  programme,  which  has
surveyed, among  others, the AEGIS-XD  and C-COSMOS fields at  100 and
$\rm  160\mu m$.   We use  the  PEP source  catalogues constructed  by
fitting the PACS PSF at  the positions of sources detected on archival
Spizter   MIPS    $24\mu   m$    data   following   the    method   of
\cite{Magnelli2009}. The $3\sigma$ depths at $\rm 100/160\mu m$ in the
AEGIS-XD  and C-COSMOS  fields  are approximately  4/8 and  5/11\,mJy,
respectively.  The  PEP source positions  were matched to  the closest
optical counterpart  using a search radius of  1.5\,arcsec.  The false
identification  rate at  the  limit $i=24$\,mag  is  3.5\%.  Total  IR
luminosities ,  $L_{IR}$, in the wavelength range  $8-1000\,\mu m$ are
determined from the PEP $\rm 100/160\mu m$ flux densities assuming the
$L_{IR}=10^{11}\,L_{\odot}$ template of \cite{Chary_Elbaz2001}.

The  prime interest  of this  paper  is the  clustering properties  of
infrared galaxies  at $z\approx1$. We therefore select  PEP sources in
the  redshift  interval  $0.6<z<1.4$.   For sources  with  photometric
redshift measurements we only use the part of the PDF that lies within
that redshift  range.  We do  not apply any  IR luminosity cut  to the
sample. At the redshift interval $0.6<z<1.4$, the PEP survey depths of
the   AEGIS-XD  and   C-COSMOS  fields   correspond  to   $L_{IR}  \ga
10^{11}\,L_{\odot}$. For  the typical stellar  mass of the  PEP far-IR
selected   galaxies   ($\rm   \approx10^{11}\,M_{\odot}$;  see   later
sections)  the luminosity limit  above corresponds  to galaxies  on or
just above the main  sequence of star-formation \citep{Santini2009} at
$z\approx1$.   Table  $\ref{table:sample}$ shows  for  each field  the
number   of  PEP   far-IR  selected   galaxies  used   for  clustering
measurements.

\subsection{The X-ray AGN samples}\label{sec_xagn}

We  use X-ray data  from the  Chandra 800\,ks  survey of  the AEGIS-XD
field (Nandra et al.  in prep) and the 2006-2007 Chandra survey of the
C-COSMOS field \citep[][]{Elvis2009}.  The Chandra observations of the
AEGIS-XD and C-COSMOS  were analysed in a homogeneous  way by applying
the   reduction  and   source  detection   methodology   described  by
\cite{Laird2009}.  The optical identification of the X-ray sources was
based        on        the        Likelihood       Ratio        method
\citep{Sutherland_and_Saunders1992}.  In  the case of  the AEGIS-XD we
used  the  IRAC-$\rm  3.6\mu  m$ selected  multi-waveband  photometric
catalogue  provided  by  the  Rainbow  Cosmological  Surveys  Database
\citep{Perez2008,  Barro2011b,  Barro2011a}.   The  identification  of
C-COSMOS  X-ray  sources  used the  \cite{Ilbert2009}  multiwavelength
photometric catalogue (Aird et al. in prep).

Extensive spectroscopic campaigns have  been carried out in the fields
of choice.   Spectroscopic redshift  measurements of X-ray  sources in
the AEGIS-XD field are primarily from the DEEP2 \citep{Newman2012} and
DEEP3     galaxy    redshift     surveys    \citep{Cooper2011_deep3_1,
Cooper2012_deep3_2}  as well as  observations carried  out at  the MMT
using the Hectospec  fibre spectrograph \citep{Coil2009}. Redshifts in
C-COSMOS  are from  the public  releases of  the  VIMOS/zCOSMOS bright
project \citep{Lilly2009} and the Magellan/IMACS observation campaigns
\citep{Trump2009}, as  well as the compilation of  redshifts for X-ray
sources presented by \cite{Brusa2010}.

For X-ray sources without spectroscopic identifications care is needed
when determining photometric redshifts  because of the contribution of
AGN light to the observed SED. Salvato et al. (2009, 2011) showed that
for  X-ray  AGN  it   is  possible  to  achieve  photometric  redshift
accuracies comparable to galaxy samples by (i) adopting priors for the
templates  used  for  each  source, (ii)  icluding  hybrid  AGN/galaxy
templates and (iii) increasing the number of photometric bands used to
sample  the  observed  SED.   For  X-ray  AGN  we  therefore  use  the
photometric redshift  PDFs estimated using  the methods of  Salvato et
al.   (2009, 2011).   These  are presented  in \cite{Salvato2011}  for
C-COSMOS and Nandra  et al. (in prep) for  the AEGIS-XD. The estimated
rms scatter of the  X-ray AGN photometric redshifts is $\sigma_{\Delta
z  / (1+z)}=0.016$  and 0.04  for the  C-COSMOS and  AEGIS-XD samples,
respectively. The corresponding outlier fraction, defined as $\Delta z
/(1+z)>0.15$, is about 6\% in both fields.

A  by-product  of  the   photometric  redshift  determination  is  the
characterisation of  the Spectral  Energy Distribution (SED)  of X-ray
AGN, e.g.  host galaxy type, level of optical extinction, level of the
AGN  component relative  to the  underlying host  galaxy.   The latter
information is  used in later  sections to identify sources  for which
the AGN  radiation likely contaminates the host  galaxy light.  Far-IR
counterparts  to X-ray  AGN are  identified  by matching  the PEP  and
optical  source catalogues in  AEGIS-XD and  C-COSMOS within  a search
radius of 1.5\,arcsec.

The  intrinsic  column density,  $N_H$,  of  individual  X-ray AGN  is
determined from the hardness  ratios between the soft (0.5-2\,keV) and
the hard (2-7\,keV) X-ray  bands assuming an intrinsic power-law X-ray
spectrum  with  index  $\Gamma=1.9$  \citep[e.g.][]{Nandra1994}.   The
derived column densities  are then used to convert  the count-rates in
the 0.5-7\,keV  band to rest-frame 2-10\,keV luminosity,  $L_X(\rm 2 -
10 \,  keV)$.  For  sources with photometric  redshift PDF,  the X-ray
luminosity  is  also  a  probability distribution  function.   In  the
clustering analysis  we use X-ray  sources with $L_X(\rm  2-10\,keV) >
10^{42}  \,  erg \,  s^{-1}$  and  redshifts  $0.6<z<1.4$.  For  X-ray
sources with photometric redshift  estimates we retain in the analysis
only  the part  of their  PDF that  corresponds to  the  limits above.
X-ray sources  in regions  that have been  masked out because  of poor
optical  photometry  (e.g.   bright   stars)  are  excluded  from  the
analysis.   Table  $\ref{table:sample}$ presents  for  each field  the
number  of  X-ray  AGN  and  the  number  of  X-ray  AGN  with  far-IR
counterparts in the PEP survey.

\subsection{Passive galaxies}\label{sec_passive}

In the following sections the  clustering of X-ray AGN and IR-selected
star-forming  galaxies   will  be   compared  to  that   of  quiescent
galaxies. The latter are selected using the rest-frame $U - V$ vs $V -
J$ (UVJ) colour-colour  diagram \citep{Williams2009, Patel2012}.  This
combination of  colours is least  sensitive to dust extinction  and is
shown   to  be  effective   in  separating   early-type,  low-specific
star-formation  rate galaxies  from  actively star-forming,  including
dust-reddened systems (Williams et al. 2009).

Figure  \ref{fig_UVJ}   plots  the   UVJ  diagram  of   galaxies  with
$18<R<24.5$\,mag and $0.6<z<1.4$ in  the AEGIS-XD and C-COSMOS fields.
The  rest-frame $U-V$  and $V-J$  colours  are estimated  by the  {\sc
kcorrect} version  4.2 routines.   For AEGIS-XD the  CFHTLS-D3 $ugriz$
and Palomar WIRC $JK$ \citep{Bundy2006} photometry is provided to {\sc
kcorrect}.   In   C-COSMOS  fluxes  in  the   CFHT  $u^\star$,  SUBARU
$Vg^+r^+i^+z^+$      \citep{Capak2007},      UKIRT      WFCAM      $J$
\citep{McCracken2010} and  CFHT WIRCAM $Ks$  \citep{Capak2007} filters
are  used.  Rest-frame $U$,  $V$ and  $J$ luminosities  are determined
from the  observed photometry in  the $r/r^{+}$, $z/z^{+}$  and $K/Ks$
bands of  AEGIS-XD/C-COSMOS, respectively. This choise of  bands is to
minimise  k-corrections.    For  sources  with   photometric  redshift
estimates the $U-V$, $V-J$  rest-frame colours are estimated separately
for  each bin  of  the photometric  redshift probability  distribution
function.   For each  of these  sources a  given ($U-V$,  $V-J$) pair,
which corresponds  to a photometric redshift  bin $z \pm  \delta z$ of
the PDF, is assinged a weight which is the probability that the source
lies in the  interval $z \pm \delta z$  (i.e.  the corresponding value
of the  photometric redshift PDF).  When constructing  the contours of
Figure  \ref{fig_UVJ}, sources  with  photometric redshifts  contribute
with different weights to different ($U-V$, $V-J$) bins.

In   Figure  \ref{fig_UVJ}  quiescent   systems  are   separated  from
star-forming (including dusty) galaxies by the selection wedge defined
by  the  relations  $U-V>1.3$,  $V-J<1.6$  and  $U-V>0.88\,(V-J)+0.69$
\citep{Williams2009}.  The specific star-formation rate of galaxies is
found  to change  rapidly across  the  wedge, at  least for  redshifts
$z\la1.5$  \citep{Williams2009}.  For the  clustering analysis  we use
sources in  the passive  region of the  UVJ diagram,  i.e.  $U-V>1.3$,
$V-J<1.6$ and  $U-V>0.88\,(V-J)+0.69$.  For galaxies  with photometric
redshift  PDFs we  use only  the part  of the  PDF  with corresponding
$U-V$, $V-J$  within the  quiescent wedge of  the UVJ  diagram.  Table
$\ref{table:sample}$  presents  for each  field  the  number of  $UVJ$
passive  galaxies used  in the  analysis. In  addition to  the optical
magnitude  ($18<R<24.5$\,mag) and  redshift  ($0.6<z<1.4$) cuts  these
samples are also selected to  include only galaxies that are bright in
absolute $J$-band magnitude, $M_J<-21.5$\,mag (see next section).

\subsection{$J$-band luminosities}\label{sec_mj}

For the interpretation and  comparison of the clustering properties of
X-ray  AGN, star-forming  and  passive galaxies  we  also explore  the
relative stellar mass distribution of their host galaxies.  We use the
$J$-band  absolute magnitude, $M_J$,  as proxy  of stellar  mass.  The
advantage  of using  luminosities  in a  near-IR  band to  approximate
stellar mass  is because the corresponding  mass--to--light ratios are
less   sensitive  to   the  star-formation   history  of   the  galaxy
\citep[e.g.][]{Bell2003}.

The  {\sc  kcorrect}  is  used  to estimate  the  rest-frame  absolute
magnitudes  (AB  system) of  extragalactic  sources  in the  2MASS-$J$
filter.   The input photometry  to {\sc  kcorrect} is  the same  as in
section \ref{sec_passive}.

To minimise k-corrections, which unavoidably depend on the adopted set
of model spectral energy  distributions, the rest-frame magnitude of a
source in the $J$ filter  is estimated from the observed photometry in
$K$-band.  At $z\approx1$  the $K$-band effective wavelength (observer
frame) is  close to that of  the $J$-band filter at  the rest-frame of
the source.   When $K$-band is not  available we use  the observed $J$
band  magnitude  to determine  $M_J$.   For  sources with  photometric
redshifts the  $M_J$ is a probability distribution  function.  We also
exclude X-ray sources  for which the SED fitting  process described in
section \ref{sec_xagn} suggests a significant AGN component that could
contaminate the host galaxy emission.  For those sources $M_J$ may not
be a proxy of stellar mass.

\begin{table*}
\caption{Details on  the samples used for  clustering measurements. We
list for  each field separately (C-COSMOS, AEGIS-XD)  and the combined
dataset (``Total'') the number  of X-ray AGN, far-IR selected galaxies
in  the PEP survey,  UVJ passive  galaxies and  X-ray AGN  with far-IR
counterparts.   The  numbers  are  sources with  either  spectroscopic
redshifts in  the range $0.6<z<1.4$ or photometric  redshifts with PDF
that  at least partially  overlaps with  the redshift  interval above.
The number of sources in  each sample with spectroscopic redshifts are
presented in  the parenthesis.   The last row  of the table  lists the
total number  of optically selected  galaxies used to  determine their
auto-correlation  function. They are  select to  have $R<24.5$  and to
satisfy  the $B-R$  vs  $R-I$ colour  cuts  defined by  Newman et  al.
(2012)      to     photometrically     pre-select      galaxies     at
$z>0.6$.}\label{table:sample} \centering \setlength{\tabcolsep}{2.5mm}

\begin{tabular}{cccc}

       \hline

{Sample} & {C-COSMOS} & {AEGIS-XD/CFHTLS-D3}  & {Total} \\


       \hline

X-ray AGN & 498 (282) & 771 (148) & 1269 (430)  \\

PEP far-IR galaxies & 578 (225) & 454 (351) & 1032 (576) \\

X-ray AGN with far-IR IDs  & 78 (50) & 103 (34) & 181 (84) \\

UVJ passive & 4087 (375)     & 817 (421) &  4883 (796) \\

galaxies &32,699 &35,991 & 68,690 \\

       \hline

\end{tabular}

\end{table*}

\begin{table*}
\caption{Power-law best-fit parameters  $r_{0}$, $\gamma$ estimated at
scales 1--10\,Mpc for the cross-correlation function between different
samples  (X-ray AGN,  far-IR  sources, X-ray  AGN with/without  far-IR
counterparts,   passive  systems)   and  optical   galaxies   and  the
auto-correlation  function of optical  galaxies.  The  inferred biases
and mean  dark matter halo masses  are also listed.   The columns are:
(1): sample  for which the  clustering properties are  estimated.  The
clustering  statistics  of the  galaxy  auto-correlation function  are
listed  in the last  row; (2):  median redshift  of each  sample; (3):
best-fit power-law  clustering scale $r_{o}$  of the cross-correlation
function with  optical galaxies  or the auto-correlation  function for
the galaxy  sample (last row); (4): best-fit  power-law index $\gamma$
of  the  cross-correlation  function  with  optical  galaxies  or  the
auto-correlation  function  for the  galaxy  sample  (last row);  (5):
reduced $\chi^2$ and  degrees of freedom for the  power-law fits to the
correlation function; (6):  cross-correlation function bias parameter,
$b_{CCF}$;    (7):   auto-correlation    function    bias   parameter,
$b_{ACF}$. For galaxies this is determined directly by measuring their
auto-correlation  function  (last row).   For  X-ray  AGN, PEP  far-IR
selected galaxies, UVJ passive galaxies, and PEP/X-ray AGN, $ b_{ACF}$
is inferred from  $b_{CCF}$ (column 6) after factoring  out the galaxy
auto-correlation function  bias listed in  the last row of  the table;
(8):   corresponding  mean   dark  matter   halo   mass.   }\centering
\setlength{\tabcolsep}{0.55mm} \setlength{\extrarowheight}{1.5mm}
\begin{tabular}{l c c c c c c c c}
\hline
 Sample & median   & $r_0$          & $\gamma$ & $\chi^2$/dof & $b_{CCF}$  & \,\,\,\,\,& $b_{ACF}$ & $\rm \log M_{DMH}$   \\
        &  redshift &($h^{-1}$\,Mpc) &          &            &      &  &      & ($h^{-1}$\,M$_\odot$)  \\
   (1)  & (2)      &   (3)          &   (4)    &  (5)   &   (6)        & & (7) & (8)   \\
\hline

& \multicolumn{4}{c}{Cross-correlation function with galaxies} & & \\

X-ray AGN    & 0.95 & $4.8^{+0.7}_{-0.6}$ & $1.9^{+0.1}_{-0.2}$ & 0.2/8   & $1.8^{+0.2}_{-0.2}$ & &$2.0^{+0.5}_{-0.4}$  & $13.0^{+0.3}_{-0.4}$  \\
PEP far-IR galaxies & 0.90 & $4.9^{+1.0}_{-0.8}$ & $1.8^{+0.1}_{-0.1}$ & 0.3/8  & $1.8^{+0.3}_{-0.2}$ & &$2.0^{+0.6}_{-0.5}$  & $13.0^{+0.4}_{-0.5}$  \\
X-ray AGN with far-IR IDs   & 0.95 & $4.9^{+1.5}_{-1.3}$ & $1.9^{+0.2}_{-0.2}$ & 0.8/8  &$1.8^{+0.5}_{-0.4}$ & & $2.1^{+1.1}_{-1.0}$  & $13.0^{+0.6}_{-1.5}$  \\ 

X-ray AGN w/out far-IR IDs  & 0.97 & $5.3^{+1.1}_{-1.1}$ &
$1.9^{+0.2}_{-0.1}$  & 1.5/8 &$1.8^{+0.2}_{-0.2}$ & & $2.1^{+0.5}_{-0.5}$  & $13.1^{+0.4}_{-0.4}$  \\


UVJ passive  & 0.84 &  $5.1^{+0.6}_{-0.9}$ & $1.6^{+0.2}_{-0.2}$  &
&$1.7^{+0.1}_{-0.2}$ & 0.3/8 & $1.8^{+0.3}_{-0.5}$  & $12.9^{+0.3}_{-0.7}$  \\

&  \multicolumn{5}{c}{Auto-correlation function of galaxies} & & \\

galaxies     & 0.90   & $4.6^{+0.6}_{-0.5}$ & $1.8^{+0.2}_{-0.1}$ & 0.2/8  &   --                & & $1.6^{+0.1}_{-0.2}$  &   --               \\                    
\hline
\label{table:fits}
\end{tabular}
\end{table*}

\begin{figure}
\begin{center}
\includegraphics[height=0.9\columnwidth]{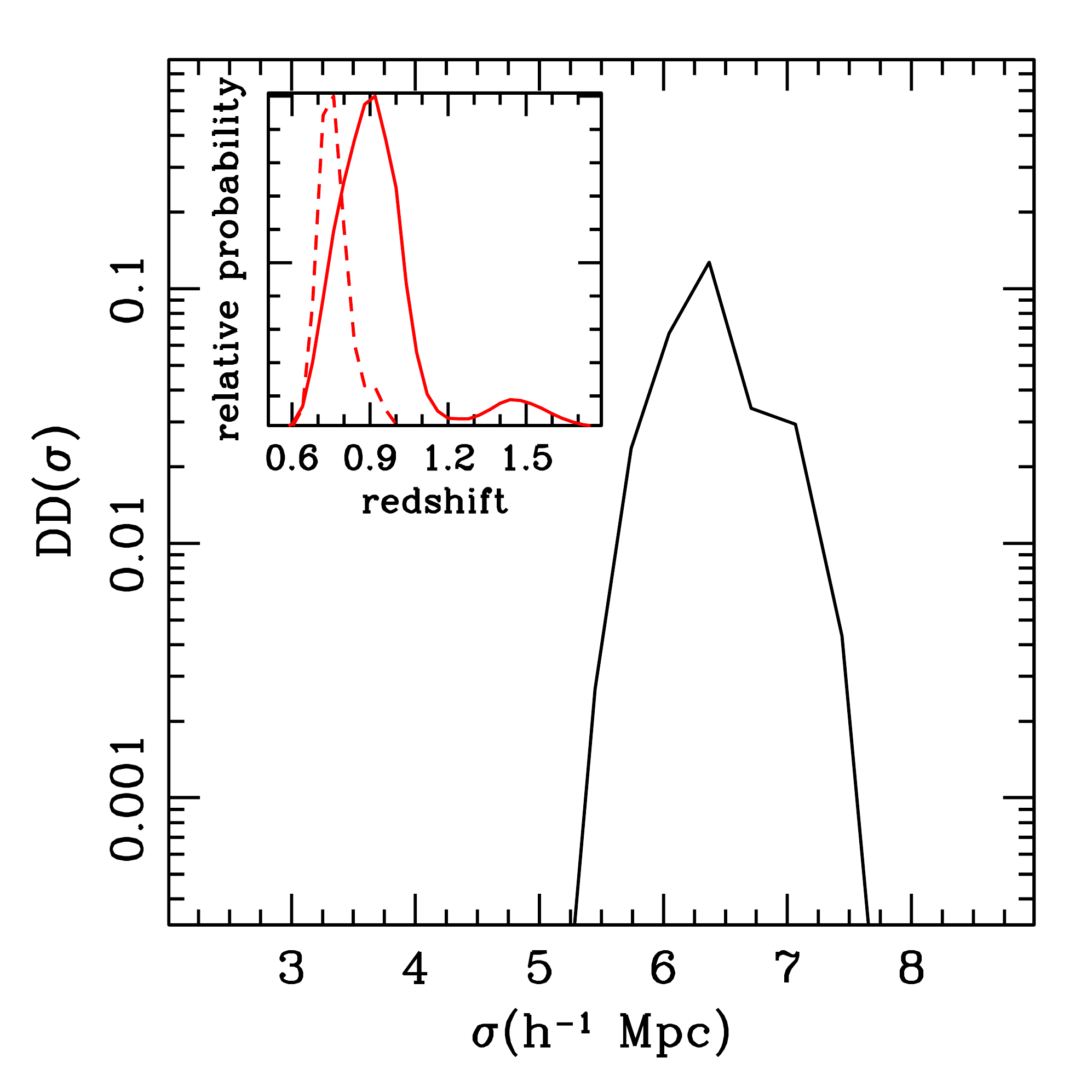}
\end{center}
\caption{Demonstration  of the  DD determination  in the  case  of the
generalised   clustering    estimator   using   photometric   redshift
information.  DD($\sigma$)  is plotted  as a function  of scale  for a
randomly chosen galaxy-galaxy pair in the AEGIS-XD field.  The inset plot
shows  the  photometric redshift  PDFs  (red  dotted  and solid  black
curves)  of  the two  galaxies.   Intuitively,  DD  in the  generalised
clustering estimator  method can be thought  of as the  product of the
convolution of the two photometric redshift PDFs.  }\label{fig_dd}
\end{figure}

\begin{figure}
\begin{center}
\includegraphics[height=0.9\columnwidth]{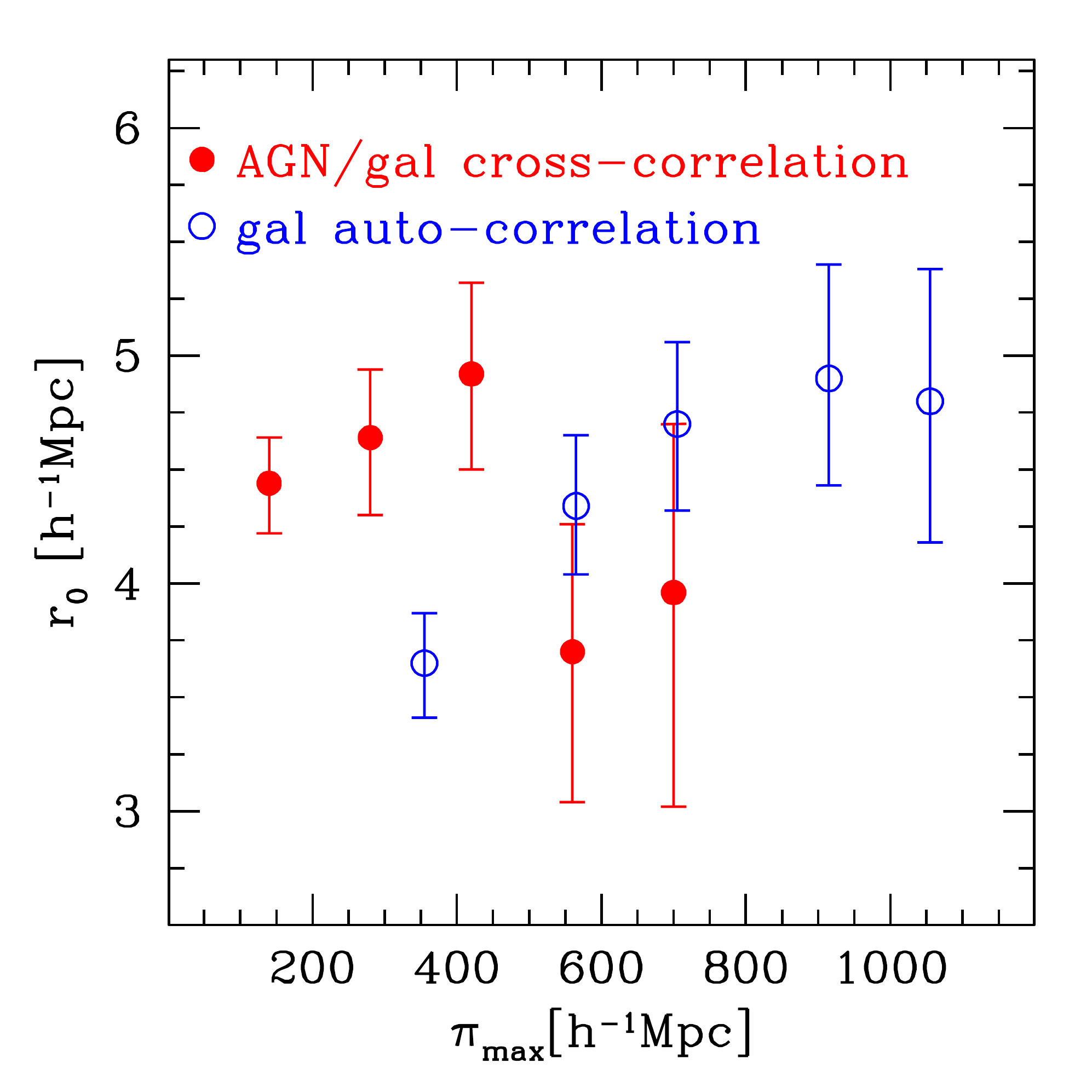}
\end{center}
\caption{Correlation  function   length,  $r_0$,  as   a  function  of
$\pi_{max}$, the maximum  scale of the integration in  equation 5. The
filled  (red)   circles  are  for   the  AGN/galaxy  cross-correlation
function.   The  open  circles  are for  the  galaxy  auto-correlation
function.   The   errors  are   jackknife.  }\label{fig_pimax}
\end{figure}

\begin{figure}
\begin{center}
\includegraphics[height=0.8\columnwidth]{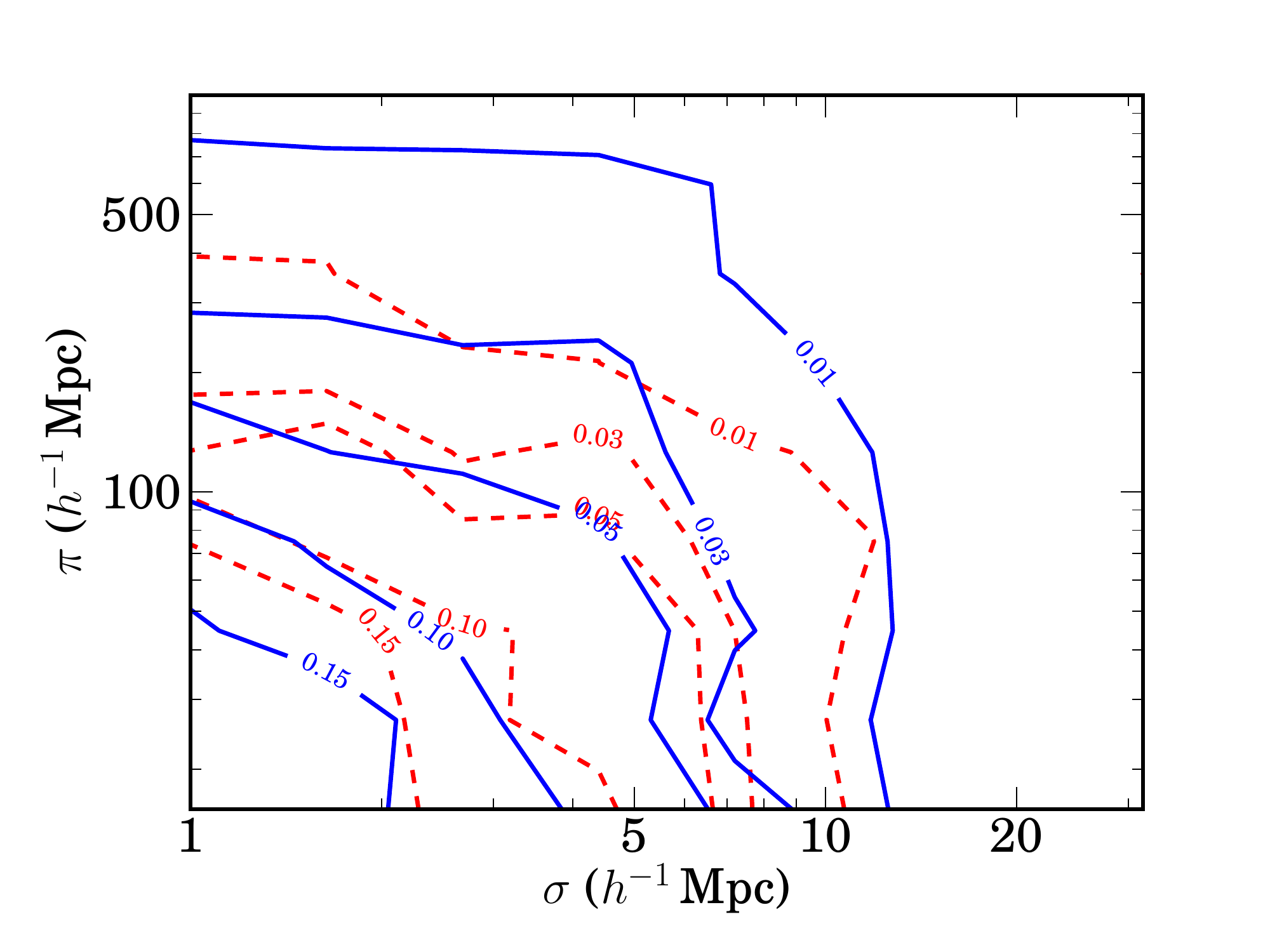}
\end{center}
\caption{Redshift-space correlation  function, $\xi(\sigma, \pi)$. The
blue  solid   contours  correspond  to   the  galaxy  auto-correlation
function. The red dashed contours are the AGN/galaxy cross-correlation
function.   The  numbers  on  the  contours  represent  the  value  of
$\xi(\sigma, \pi)$. The elongations along the $\pi$ direction (y-axis)
are   because  of   peculiar  velocities   and   photometric  redshift
uncertainties.   The  contours  show   that  the  integration  of  the
$\xi(\sigma,  \pi)$ to  $\pi_{max}=450$, 700\,Mpc  for  the AGN/galaxy
cross-correlation   and    the   galaxy   auto-correlation   functions
respectively, includes the bulk of the clustering signal.  }\label{fig_xi}
\end{figure}

\section{Methodology:  generalised clustering estimator for photometric redshift samples}\label{section:method}


Next we present the equations  used to determine the clustering of any
extragalactic  population   such  as   AGN  or  galaxies.    Both  the
auto-correlation and the cross-correlation functions are special cases
of  the  2-point  statistics   of  the  AGN  and  galaxy  populations.
Therefore, they are both defined by the same basic equations.  In this
section   the  term   correlation  function   refers  to   either  the
auto-correlation or the cross-correlation functions. When necessary we
will  differentiate  between   the  two  quantities.   The  real-space
correlation  function,  $\xi(r)$, can  be  estimated  by the  relation
\citep{Davis_Peebles1983}

\begin{equation}
\label{eqn:xi} \xi(r) =\frac{DD(r)}{DR(r)}-1,
\end{equation}

\noindent where DD(r) are the data-data pairs at separation $r$. DR(r)
are  the AGN-random pairs  (cross-correlation) or  galaxy-random pairs
(galaxy auto-correlation function) at separation $r$.  Both DD and
DR  in equation  \ref{eqn:xi} are  normalised  appropriately.  Random
catalogues  are  produced by  randomising  the  position of  galaxies,
taking  into account  the sample  selection function,  i.e.  magnitude
limit,  field  boundaries, masked  regions.   We  choose to  construct
random catalogues for the galaxy (tracer) sample.  This is because the
spatial selection function of the  X-ray AGN in particular, is complex
and varies across  the field of view of  individual Chandra pointings.
These  variations might  introduce systematics  into  the calculations
unless they are quantified to a high degree of accuracy.

The distance $r$ can be  decomposed into separations along the line of
sight, $\pi$,  and across  the line of  sight, $\sigma$. If  $s_1$ and
$s_2$  are   the  distances   of  two  objects   1,  2,   measured  in
redshift-space, and $\theta$ the angular separation between them, then
$\sigma$ and $\pi$ are defined as

\begin{equation} 
\pi=(s_2-s_1), $ along the line-of-sight$,
\end{equation}

\begin{equation}
\sigma=\frac{(s_2+s_1)}{2}\theta , $ across the line-of-sight$.
\end{equation}

\noindent The correlation function in redshift-space is then estimated
as

\begin{equation}
\xi(\sigma,\pi) =\frac{DD(\sigma,\pi)}{DR(\sigma,\pi)}-1.
\label{eqn:wtheta}
\end{equation}

\noindent  In the  classic approach  of estimating  the redshift-space
correlation  function,  when   accurate  spectroscopic  redshifts  are
available,  each data-data  pair with  $\sigma$, $\pi$  separations is
incremented by  one, i.e.  $DD(\sigma,\pi)=DD(\sigma,\pi)+1$.   If the
redshift determinations are uncertain, i.e. photometric redshifts, the
above relation  can be generalised to include  those uncertainties, in
the form of Probability Distribution Functions (PDF).

Suppose  two data  points, D1 and  D2, which are  associated with
photometric redshift probability  distribution functions $\rm PDF_{1}$
and  $\rm  PDF_{2}$, respectively.   We  assume  that  these PDFs  are
estimated at  discrete photometric  redshift data points  with binsize
$\delta  z$.   The  value $f_i$  of  a  PDF  at  the  bin $i$  is  the
probability that the source lies in the redshift range $z_i \pm \delta
z/2$.  Let  us then  assume a probability  $f_{i,1}$, drawn  from $\rm
PDF_{1}$, that D1  lies in the redshift slice  $z_i\pm\delta z/2$. The
probability  of D2  lying at  separations $\sigma$,  $\pi$ from  D1 at
$z_i$ can then be estimated  from $\rm PDF_{2}$, e.g.  $f_{i,2}$.  The
number of data-data pairs  $DD(\sigma,\pi)$ is then incremented by the
product     $f_{i,1}\,f_{i,2}$,     instead     of     unity,     i.e.
$DD(\sigma,\pi)=DD(\sigma,\pi)+f_{i,1}\,f_{i,2}$.    In  this  picture
DD(r)  corresponds  to a  probability  distribution function.   Figure
\ref{fig_dd}  shows  an  example  of  a  single  DD  estimated  for  a
particular  galaxy-galaxy  pair  in   the  AEGIS-XD  field.   In  this
particular case the photometric redshift  PDF has a binsize of $\delta
z  = 0.01$.  The  $DR(\sigma,\pi)$ pairs  are estimated  following the
same procedure. Each random point is assigned the photometric redshift
PDF  of one  of the  observed galaxies  in the  sample.   The redshift
distribution  of  random  points  is  therefore  similar  to  that  of
galaxies.

When  the  correlation function  is  measured  in redshift-space,  the
clustering  is  affected at  small  scales  by  the peculiar  velocity
component  of extragalactic  sources along  the line  of sight  and by
dynamical infall of matter into higher density regions. In the case of
the    generalised   clustering    estimator    photometric   redshift
uncertainties  also  have  an   impact  on  the  radial  component  of
$\xi(\sigma,\pi)$. These  effects can be removed  by integrating along
the line of sight, $\pi$, to calculate the projected cross-correlation
function

\begin{equation} w_p(\sigma)=2\int_0^{\pi_{max}} \xi(\sigma,\pi)d\pi.
\label{eqn:wp}
\end{equation}

\noindent The maximum scale of  the integration is a trade-off between
underestimating the clustering amplitude, if $\pi_{max}$ is too small,
and  low signal-to-noise  ratio,  if $\pi_{max}$  is  too large.   The
optimum $\pi_{max}$  value can be  determined by either  (i) measuring
the projected correlation function  for different $\pi_{max}$ and then
adopting  the value at  which the  amplitude of  the cross-correlation
function appears to  level off or (ii) by  inspecting how $\xi(\sigma,
\pi)$ is distributed in $\sigma$, $\pi$ space and then determining the
maximum $\pi$ value that includes most of the clustering signal.

The uncertainties  of the  correlation function at  a given  scale are
estimated  using the  Jackknife  methodology.  The  survey fields  are
divided into a total  of $N_{JK}$ sections.  The projected correlation
function is re-estimated $N_{JK}$ times by excluding in each trial one
of the  sections. These  measurements are then  used to  determine the
covariance matrix,  which quantifies the level  of correlation between
different different  bins of $w_p(\sigma)$ \citep[e.g.][]{Krumpe2010}.
During this process it was  found that the $w_p(\sigma)$ measured from
individual Jackknife sub-samples do not follow the Normal distribution
but are skewed by outliers. This effect is stronger in the case of the
generalised  clustering  estimation method.   We  therefore choose  to
represent the  uncertainties of the  correlation function by  the 16th
and  84th percentiles  of the  distribution of  $w_p(\sigma)$ measured
from the $N_{JK}$ Jackknife sections.

The  bias parameter  for  a  given  extragalactic population  is
estimated from the rms fluctuations of the density distribution over a
sphere with  a comoving radius of  $8\,h^{-1}$\,Mpc ($\sigma_8$) under
the  assumption  that the  correlation  function  follows a  power-law
\cite[e.g.][]{Mountrichas2013}

\begin{equation} \sigma_{8}^2=J_2(\gamma)\left( \frac{r_0}{8h^{-1}Mpc}
\right)^\gamma,
\end{equation}

\noindent where

\begin{equation}
J_2(\gamma)=\frac{72}{(3-\gamma)(4-\gamma)(6-\gamma)2^\gamma},
\end{equation}

\noindent and  $\gamma$, $r_{0}$  are the slope  and amplitude  of the
power-law  form  of  the   correlation  function.  The  bias  is  then
calculated by the relation

\begin{equation} b = \frac{\sigma_{8}}{\sigma_8(z)}.
\label{eqn:bs8}
\end{equation}

\noindent  where $\sigma_8(z)$  is the  rms fluctuations  of  the dark
matter  density field  within an  $8\,h^{-1}$\,Mpc sphere  at redshift
$z$.   We account  for  the non-Gaussian  errors  of $w_p(\sigma)$  by
determining  separately  for  each  Jackknife region  the  correlation
function  power-law parameters  (slope, $\gamma$;  amplitude, $r_{0}$)
and the corresponding bias at scales 1-10\,Mpc.  The errors of each of
those parameters are then represented  by 16th and 84th percentiles of
the distribution of the  $N_{JK}$ measurements.  The covariance matrix
is used  indirectly in the  error estimation process to  determine the
best-fit power-law parameters for each Jackknife region.

We adopt  the ellipsoidal collapse  model of \cite{Sheth2001}  and the
analytical approximations of  \cite{vandenBosch2002} to infer the mean
dark matter halo mass of an extragalactic population (AGN or galaxies)
from the  measured bias parameter.   This calculation assumes  that on
large scales the bias depends only on halo mass.

Appendix A demonstrates the  performance of the generalised clustering
estimator that uses photometric redshifts PDFs. The results using this
method   are   compared   with   clustering  measurements   based   on
spectroscopic samples only.  It is  shown that the method described in
this  section  can  recover  the clustering  signal  of  extragalactic
populations   even  if  no   spectroscopic  redshift   information  is
available. Large photometric redshift samples are required however, at
least  10  times  larger  than  spectroscopic  ones,  to  recover  the
clustering signal at the same level of accuracy.

\begin{figure*}
\begin{center}
\includegraphics[height=0.9\columnwidth]{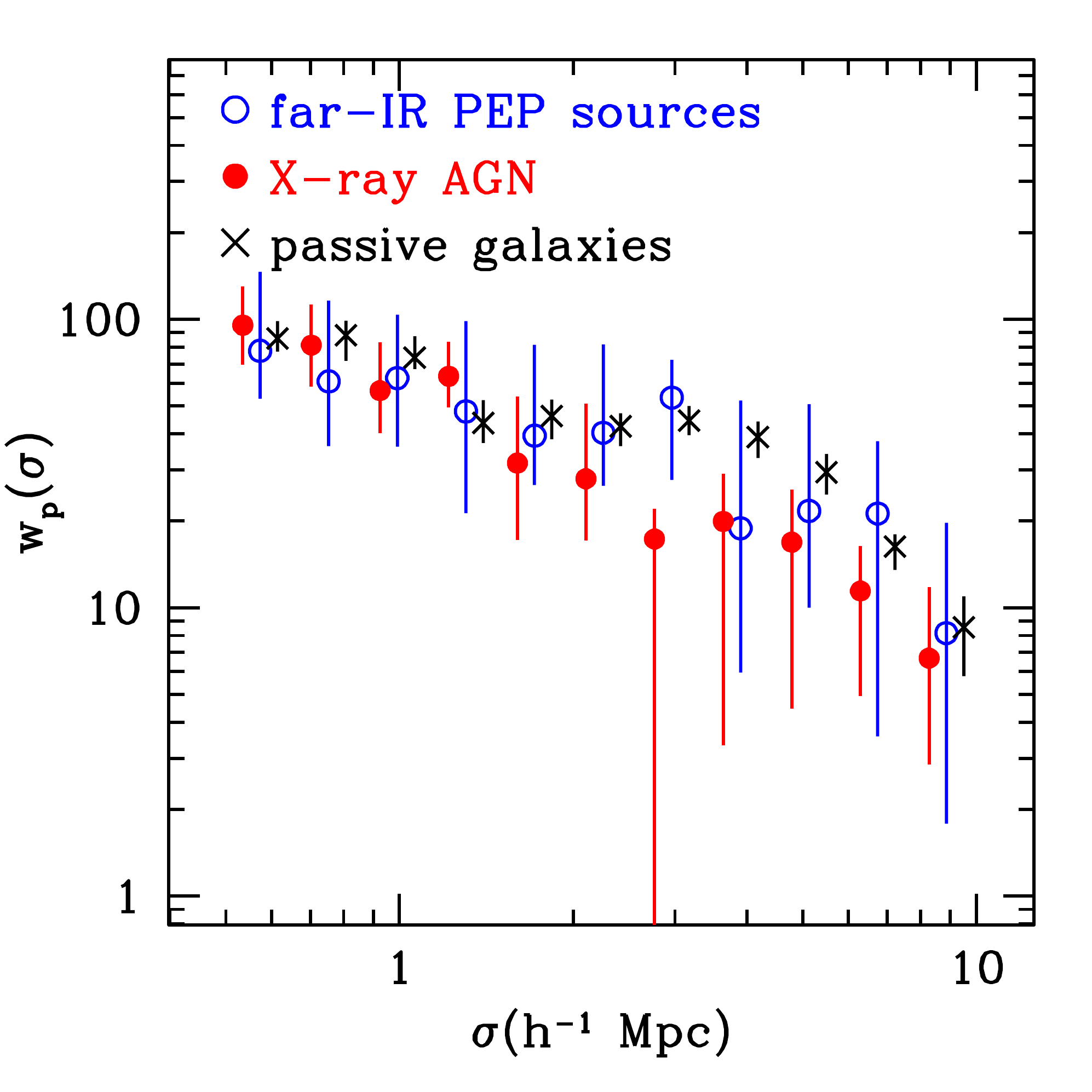}
\includegraphics[height=0.9\columnwidth]{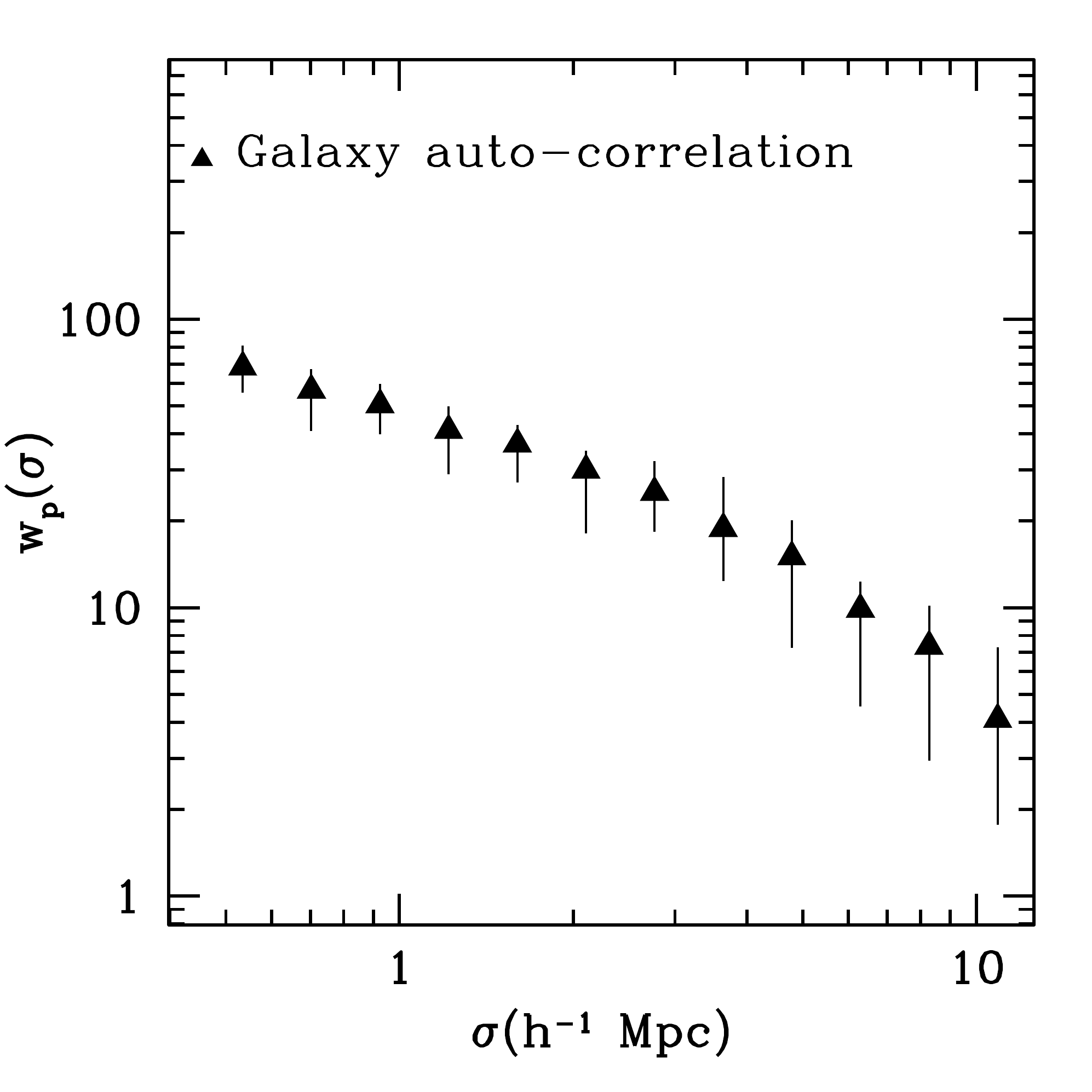}
\end{center}
\caption{{\bf Left:}  Cross-correlation function with  galaxies of (i)
X-ray AGN  (solid red circles),  (ii) IR sources (empty  blue circles)
and  (iii) $UVJ$  passive galaxies  (black crosses),  in  the combined
AEGIS-XD and C-COSMOS fields.  The errorbars are Jackknife.  The solid
red symbols are offset in the horizontal direction by $\log = \pm0.03$
for  clarity.   {\bf  Right:}  Auto-correlation function  of  galaxies
($R<24.5$\,mag)  in the  combined AEGIS-XD  and C-COSMOS  fields.  The
uncertainties of individual  datapoints are estimated from $N_{JK}=16$
Jackknife sub-regions.}\label{fig_final_cross}
\end{figure*}

\section{Results}\label{sec_res}

In this section  we estimate and compare the  clustering properties of
(i) X-ray AGN, (ii) far-IR selected sources detected in the PEP survey
and  (iii)  passive  galaxies   selected  by  their  $U-V$  and  $V-J$
rest-frame        colours       \citep{Williams2009,       Patel2012}.
Differences/similarities  of  the  large  scale environment  of  those
samples can provide  clues on the association between  AGN activity as
traced by X-rays and the level of star-formation in galaxies.  In this
comparison one  should also  account for possible  covariances between
environment   and   galaxy   properties   other   than   instantaneous
star-formation rate.  One galaxy  parameter that is known to correlate
with      large-scale       environment      is      stellar      mass
\citep[e.g.][]{Mostek2012}.   Therefore in  the interpretation  of the
clustering properties of the samples above we also include information
on  the stellar mass  distribution of  the underlying  galaxies. Also,
X-ray AGN and far-IR  sources have similar redshift distributions that
both  peak  at $z\approx0.9$  (see  Table \ref{table:fits}).   Passive
galaxies however, because of their  red SEDs, have a distribution that
peaks    at   somewhat   lower    redshift,   $z=0.84$    (see   Table
\ref{table:fits}).     In the  following  calculations and  unless
otherwise  stated,  we  use   both  photomertric  redshifts  PDFs  and
spectroscopic  redshifts,  when   available,  for  X-ray  AGN,  far-IR
selected  sources  detected  in  the  PEP  survey  and  UVJ  quiescent
galaxies.  For  the galaxy auto-correlation  function only photometric
redshift PDFs are used. 

In the clustering calculations that use photometric redshift PDFs
we  adopt  $\pi_{max}   =  450$  and  700\,Mpc  for   the  cross-  and
auto-correlation functions, respectively. These values are larger than
what is typically adopted in clustering studies that use spectroscopic
samples    only,     e.g.     $\pi_{max}    =     40    -    100$\,Mpc
\citep[e.g.][]{Coil2009, Krumpe2010,  Mountrichas2012}. The difference
is because of  the larger uncertainties of the  redshifts measured via
photometric methods.   Figure \ref{fig_pimax} demonstrates  the choice
of   $\pi_{max}$  for  the   photometric  redshift   subsamples.   The
AGN/galaxy cross-correlation and  the galaxy auto-correlation function
amplitudes are determined for  different $\pi_{max}$ values.  For each
sample  we  choose the  $\pi_{max}$  at  which  the clustering  signal
appears to level off.   An alternative method to determine $\pi_{max}$
is  shown  in  Figure   \ref{fig_xi}.   It  plots  the  redshift-space
correlation   function  $\xi(\sigma,   \pi)$  for   both   the  galaxy
auto-correlation   function  and   the   AGN/galaxy  cross-correlation
function.     That   figure    shows   that    integration    of   the
auto-/cross-correlation  function  to   $\pi_{max}  \approx  450$  and
700\,Mpc  includes  all  the  clustering  signal.   These  values  are
consistent  with those  determined from  Figure  \ref{fig_pimax}.  The
value  $\pi_{max}  \approx  450$\,Mpc  is  also  appropriate  for  the
cross-correlation  function of  galaxies with  either  far-IR selected
sources  or UVJ passive  systems.  The  $\pi_{max}$ behavior  of those
samples are  not plotted in Figures  \ref{fig_pimax}, \ref{fig_xi} for
clarity. 

The  difference  between  the   $\pi_{max}$  for  the  AGN/galaxy  and
galaxy/galaxy correlation  functions is related to  differences in the
construction  of the photometric  redshift PDFs  of AGN  and galaxies.
Aliases among different templates is a known limitation in photometric
redshift    estimates,   particularly    in   the    case    of   AGN.
\cite{Salvato2009,  Salvato2011} manage to  minimise this  problem for
AGN  by applying  priors  based on  source  properties, e.g.   optical
extent, X-ray flux.   Depending on those priors only  subsets of their
full template  library are used to estimate  the photometric redshifts
and the  corresponding PDFs  for individual sources.   As a  result of
narrowing down  the template space  the photometric redshift  PDFs for
the  AGN  sample are  typically  narrower  than  those of  the  galaxy
population.

Finally, for passive  galaxies we also present clustering results
using  the classic  cross-correlation  and auto-correlation  functions
based on  spectroscopy only (see  Section \ref{sec_res_passive}).  For
this calculation we use $\pi_{max} = 50$\,Mpc.

\subsection{Clustering of X-ray AGN and far-IR galaxies}\label{sec_res_xray}

The cross-correlation  function of  X-ray and far-IR  selected sources
with  galaxies is  determined by  applying the  generalised clustering
estimator  methodology  presented  in  the previous  sections  to  the
AEGIS-XD and C-COSMOS fields.   The number of sources used in the
calculation  are listed  in Table  \ref{table:sample}.   The combined
correlation function is determined by equation \ref{eqn:wtheta}, where
DD and  DR in this case are  the sum of the  data-data and data-random
pairs,  respectively,  in  the   two  fields.   Motivated  by  Figures
\ref{fig_pimax}, \ref{fig_xi} the $\pi_{max}$ is set to 450\,Mpc.  For
the interpretation  of the cross-correlation function  we also measure
the  auto-correlation  function of  the  galaxy  sample following  the
methodology of  section \ref{section:method}.  In  this calculation we
use $\pi_{max}=700$\,Mpc (see Figures \ref{fig_pimax}, \ref{fig_xi}).

The uncertainties  of the  correlation function at  a given  scale are
estimated  by dividing  the two  fields  into a  total of  $N_{JK}=16$
sections  (8  for each  of  the  two  survey fields).   The  projected
cross-correlation  functions with  galaxies  of X-ray  AGN and  far-IR
selected    star-forming     galaxies    are    shown     in    Figure
\ref{fig_final_cross}-left.   The   amplitude,  $r_0$,  and  exponent,
$\gamma$, of the best-fit  power-law at scales 1-10\,Mpc are presented
in  Table \ref{table:fits}.   The relative  cross-correlation function
bias of the two populations is $b_{AGN}/b_{IR}=0.97 \pm 0.20$.  Within
the errors the two populations have consistent clustering properties.

We further estimate the mean dark  matter halo of X-ray AGN and far-IR
sources  detected  in  the  PEP  survey.   This  calculation  requires
knowledge of  the galaxy auto-correlation function, which  can then be
factored  out   of  the  cross-correlation  function.     In  this
calculation it is  assumed that $b^2_{CCF}=b_{sample}\,b_{gal}$, where
$b_{sample}$  is  the bias  of  either  X-ray  AGN or  far-IR  sources
detected in the PEP survey, $b_{gal}$ is the galaxy bias inferred from
their auto-correlation  function and  $b_{CCF}$ is the  bias estimated
from       the        cross-correlation       function.        Figure
\ref{fig_final_cross}-right   plots  the   projected  auto-correlation
function of  galaxies.  This  is also fit  with a single  power-law at
scales   1-10\,Mpc.   The  resulting   best-fit  parameters   and  the
corresponding galaxy bias are listed in Table \ref{table:fits}.  Using
these results we estimate a mean  dark matter halo mass of about $\log
M/(M_{\odot}\,h^{-1})\approx13.0$  for  both   X-ray  AGN  and  far-IR
sources respectively.  For the latter population the above dark matter
halo    mass    is     consistent    with    recent    estimates    by
\cite{Magliocchetti2011,  Magliocchetti2013}.  Using PEP  survey data,
they  infer   a  mimimum  halo   mass  for  their  sources   of  $\log
M/M_{\odot}\approx12.0-12.4$.   The   apparent  discrepancy  with  our
results is  related to  the method adopted  to infer dark  matter halo
masses   from  the  measured   correlation  function.    Applying  our
methodology  (see section \ref{section:method})  to the  amplitude and
power-law index of the  real-space correlation functions determined by
\cite{Magliocchetti2011, Magliocchetti2013},  we estimate a  mean halo
mass $\log  M / (M_{\odot}\,h^{-1})\approx13.0$, i.e.   similar to the
value we infer via the cross-correlation with galaxies.

For  completeness, we also estimate the  clustering properties of
the sub-samples of X-ray AGN with and without IR counterparts. Because
of  the  small   number  of  sources  in  the   former  subsample  the
corresponding clustering signal is noisy.  The relative bias of X-ray
AGN with and without IR  counterparts is $b_{rel}=1.03 \pm 0.32$.  The
corresponding  dark  matter  halos  are $\log  M/  (M_{\odot}\,h^{-1})
=13.0^{+0.6}_{-1.5}$     and    $13.1^{+0.4}_{-0.4}$     (see    Table
\ref{table:fits}).  Within the errors  we find no  differences in
the clustering properties of the two sub-populations.

Figure  \ref{fig_mj_hist}  plots the  distribution  of  X-ray AGN  and
far-IR   selected   star-forming   galaxies   in   $J$-band   absolute
magnitude. There  is considerable overlap between  the two populations
thereby,   indicating  similar   stellar   mass  distributions.    For
reference, $M_J=-21.5$  and $-23.0$\,mag correspond  to stellar masses
$\log M_{\star}/M_{\odot}\approx10.5$  and 11.0 respectively, assuming
the $J$-band  mass--to--light ratios  of \cite{Bell2001} and  a galaxy
rest-frame colour  $V-J=1$\,mag (AB  system).  It is  therefore likely
that the similar mean dark matter  halo masses of X-ray AGN and far-IR
selected  star-forming  galaxies is  a  consequence  of their  similar
stellar mass distributions approximated by $M_J$.

\subsection{Clustering of quiescent galaxies}\label{sec_res_passive}

We  also  compare the  above  results  with  the clustering  of  $UVJ$
selected  passive galaxies  in the  redshift interval  $0.6<z<1.4$. In
this comparison we also attempt to  have a control on the stellar mass
of the  passive galaxy sample. As  in the previous section  we use the
$J$-band  luminosity as  proxy  of stellar  mass  and apply  a cut  of
$M_J<-21.5$.   The  corresponding $M_J$  distribution  of the  passive
sample is  compared in Figure  \ref{fig_mj_hist} to that of  X-ray AGN
and far-IR  selected galaxies.  All  three samples have  similar $M_J$
distributions.

The  clustering of  the passive  galaxy sample  ($U-V>1.3$, $V-J<1.6$,
$U-V>0.88\,(V-J)+0.69$,  $M_J<-21.5$,   $0.6<z<1.4$; see  Table
\ref{table:sample})  is estimated  via their  cross-correlation with
the  overall galaxy population.   The results  are presented  in Table
\ref{table:fits}.    We   estimate  a   bias   for   this  sample   of
$b=1.8^{+0.3}_{-0.5}$ and  an average dark  matter halo mass  of $\log
M/(M_{\odot}\,h^{-1})=12.9^{+0.3}_{-0.7}$.  The errors in the inferred
dark matter halo  mass are large and pose  a limitation when comparing
to X-ray AGN and far-IR selected galaxies.

Therefore,  for this  particular application  we turn  to  the classic
clustering  estimator that  uses spectroscopic  redshifts measurements
only.  We  exploit the extensive  and homogeneous spectroscopy  in the
AEGIS-XD field  to infer  the clustering properties  of the  subset of
spectroscopically confirmed $UVJ$ passive  galaxies (total of 421; see
Table  \ref{table:sample}  and  section \ref{sec_passive})  via  their
cross-correlation with the overall spectroscopic galaxy sample in that
field.  The C-COSMOS field is not used in this exercise because of the
sparser and  significantly more complex sampling  of the spectroscopic
galaxy     sample      \citep[e.g.][]{delaTorre2011}.      For     the
spectroscopically confirmed UVJ passive  galaxies in AEGIS-XD field we
measure        a         cross-correlation        function        bias
$b_{CCF}=1.7^{+0.1}_{-0.1}$. We then infer an auto-correlation bias of
$b=1.9^{+0.2}_{-0.2}$ and  an average dark  matter halo mass  of $\log
M/(M_{\odot}\,h^{-1})=13.1^{+0.1}_{-0.2}$.  Within  the errors this is
similar to the mean dark matter halo masses measured for X-ray AGN and
far-IR selected star-forming galaxies.

\section{Discussion}

\subsection{Clustering measurements with photometric redshifts}

In  this paper we  present a  novel method  to estimate  the projected
correlation  function   of  extragalactic  sources,   which  is  least
dependent on  spectroscopic redshift  measurements.  It is  shown that
photometric   redshift    probability   distribution   functions   can
effectively   substitute  spectroscopic   redshifts  to   recover  the
clustering  signal  of extragalactic  populations.   This approach  is
geared toward large samples. The  loss of accuracy because of the lack
of  spectroscopy  can  be  balanced  by increasing  the  size  of  the
population used to  measure the clustering.  It is  found for example,
that  photometric redshift  samples with  sizes of  at least  10 times
larger than spectroscopic ones  are required to recover the clustering
of  galaxies  (auto-correlation  function)   at  a  similar  level  of
accuracy.   The  proposed methodology  is  well  suited to  clustering
investigations  using  future large  X-ray  AGN  surveys  such as  the
eROSITA  All Sky  Survey  \citep[eRASS;][]{Merloni2012, Kolodzig2013}.
Follow-up  optical   spectroscopy  for  such  large   AGN  samples  is
challenging and  may suffer from  incomplete or patchy  coverage.  The
determination of AGN photometric redshifts in wide-area surveys is not
straightforward  either. A good  coverage of  the SED  from UV  to the
near-IR  is needed  to  resolve  aliases and  reduce  outliers in  the
photometric   redshift    determinations   \citep{Salvato2011}.    The
inclusion  of  intermediate/narrow-band  filters  is  also  desirable,
particularly  for  AGN,  which  often exhibit  strong  emission  lines
(Salvato et al. 2009,  2011).  Nevertheless, compared to spectroscopy,
homogeneous   and  well   calibrated   multi-waveband  photometry   is
relatively easier to obtain over large sky areas.

The errors  of the AGN-galaxy cross-correlation  function are expected
to scale roughly as the square  root of the number of AGN/galaxy pairs
at a  given scale, $\sqrt{N_{pairs}}$. To the  first approximation the
number of AGN/galaxy pairs is proportional to the number of AGN within
the  survey area,  $N_{AGN}$,  and the  surface  density of  galaxies,
$n_{GAL}$.  Therefore the  AGN/galaxy cross-correlation bias scales as
$\sqrt{n_{GAL} \times  N_{AGN}}$.  In  this calculation the  impact of
sample variance  in the error  budget \citep[see also][]{Kolodzig2013}
is  assumed   to  be  small.    We  can  therefore   make  approximate
calculations  on  the level  of  uncertainty  in  the AGN/galaxy  bias
parameter one  should expect  for different survey  setups. We  use as
starting point  for the calculations  the AGN/galaxy cross-correlation
bias parameter  estimated in the AEGIS-XD. That  calculations uses AGN
with photometric redshifts ($N_{AGN}=771$;  see Table 1) and CFHTLS-D3
galaxies to $R=24.1$\,mag in the  range $z=0.6-1.4$ with a sky density
of about  $\rm n_{gal}=23,000\,deg^{-2}$.   The relative error  of the
AGN/galaxy bias  $\delta b/  b$ we estimate  for that sample  is about
0.15.    The   eROSITA  will   reach   a   flux   limit  of   $f_X(\rm
0.5-2\,keV)\approx10^{-14}  \, erg  \,  s^{-1} \,  cm^{-2}$ after  the
completion of  the 4-year all-sky  survey plan.  Folding  the observed
numbers  counts in  the 0.5-2\,keV  band \citep{Georgakakis2008_sense}
with the expected sensitivity of  the eROSITA 4-year all sky survey we
estimate an AGN surface density  of about $\rm 40\,deg^{-2}$.  We then
assume the  $\rm 5000\,deg^2$  area of the  Dark Energy  Survey (DES),
which   will    yield   photometric   redshifts    for   galaxies   to
$R\approx24$\,mag,  i.e.  similar   magnitude  limit  adopted  in  the
CFHTLS-D3 field.   For this setup  we estimate an  AGN/galaxy relative
bias uncertainty $\delta b/ b=0.01$.   We caution that this is a lower
limit to the error budget because (i) the impact of cosmic variance is
ignored  and (ii)  the final  uncertainty  in the  inferred AGN  bias,
$b_{AGN}$,  also   depends  on  the   accuracy  of  the   galaxy  bias
determination and  (iii) the calculation assumes  that the photometric
redshift accuracy for X-ray AGN is the same as in the AEGIS-XD field.

With  respect to  the latter  point, it  should be  expected  that the
predictions above  depend on the accuracy of  the photometric redshift
determinations of both galaxies and AGN.  In the case of wide-area and
shallow  X-ray   surveys  in  particular,  such  as   eRASS,  this  is
potentially  a   serious  limitation  to   clustering  investigations.
Firstly,  such surveys  include a  large faction  of luminous  AGN for
which  accurate  photometric  redshifts  are challenging  to  estimate
because  light  from both  the  host  galaxy  and the  central  engine
contribute to the observed SED (e.g.  Salvato et al.  2011).  Secondly
multi-band  photometry, like  that available  in the  C-COSMOS (nearly
30-bands including narrow filters,  Salvato et al.  2011) and AEGIS-XD
(up to  35 bands, Nandra  et al.  in  prep) fields, is hard  to obtain
over thousands  of square degrees  on the sky.  We  parameterise X-ray
AGN photometric redshift errors  by assuming that the quantity $\Delta
z   /  (1+z)$   is   distributed  as   a   Gaussian  with   dispersion
$\sigma_{\Delta z  / (1+z)}$ (e.g.   Salvato et al.  2009,  2011). For
reference   the  C-COSMOS   and  AEGIS-XD   X-ray  AGN   samples  have
$\sigma_{\Delta z/(1+z)}\approx0.016$ and 0.04, respectively. We limit
the AEGIS-XD X-ray AGN  sample to sources with spectroscopic redshifts
only  and convolve them  with a  Gaussian filter  with $\sigma_{\Delta
z/(1+z)}$ in  the range 0.01 to  0.08.  For each  convolved sample the
cross-correlation function with CFHTLS-D3 galaxies is determined.  For
$\sigma_{\Delta    z/(1+z)}\la0.04$   we   estimate    an   AGN/galaxy
cross-correlation  bias  of  $\approx2.0\pm0.3$, consistent  with  the
estimates   presented   in   section  \ref{sec_validate}   and   Table
\ref{table:fits}.  For  larger photometric uncertainties  however, the
methodology of  estimating clustering using  photometric redshift PDFs
breaks down.  In  this case we find it is not  possible to determine a
stable  $\pi_{max}$ for  the AGN/galaxy  cross-correlation,  e.g.  the
clustering amplitude does not level off with increasing $\pi_{max}$ as
in Figure \ref{fig_pimax}.

The  accuracy $\sigma_{\Delta  z/(1+z)}\approx0.04$ is  challenging to
obtain, particularly for bright AGN  samples, when only a small number
of optical  or optical/near-IR bands are available  (e.g.  see Figures
13 of Salvato et al. 2011). Alternatively one may improve the accuracy
of the galaxies' photometric redshifts that are used in the estimation
of the cross-correlation function.  We explore this possibility in the
C-COSMOS  field,  where  $\sigma_{\Delta z/(1+z)}=0.01$  for  galaxies
brighter  than   $i=24$\,mag  (Ilbert   et  al.   2009)   compared  to
$\sigma_z/(1+z)=0.03$ in the AEGIS-XD/CFHTLS-D3 (Coupon et al.  2009).
We produce AGN samples with photometric redshifts of variable accuracy
by  convolving the  spectroscopic redshifts  of X-ray  sources  in the
C-COSMOS  survey  with  a  Gaussian filter  of  width  $\sigma_{\Delta
z/(1+z)}$. The  latter parameter varied  for each sample in  the range
0.01-0.08.   The results  are  plotted in  Figure \ref{fig_sim}.   The
clustering  of X-ray  AGN  via their  cross-correlation function  with
galaxies    can    be     recovered    even    when    $\sigma_{\Delta
z/(1+z)}=0.08$. Therefore, a  galaxy sample with photometric redshifts
accurate  at   the  $\sigma_{\Delta  z/(1+z)}\approx0.01$   level  can
compensate for  larger photometric  redshift uncertainties of  the AGN
sample.   It is  also  interesting that  in  Figure \ref{fig_sim}  the
uncertainties of the  inferred bias are nearly independent  of the AGN
photometric redshifts dispersion, $\sigma_{\Delta z/(1+z)}$. The total
number  of AGN  used to  measure the  cross-correlation  function with
galaxies is the  factor that predominantly affects the  level of error
in the estimated bias.

Finally, it can also be shown that the determination of the clustering
properties of  eROSITA AGN via  the auto-correlation function  is less
competitive than  the AGN/galaxy cross-correlation  function.  In this
case the errors should scale as $\sqrt{n_{AGN} \times N_{AGN}}$, where
$n_{AGN}$  is  the   sky  density  of  AGN  sources.    To  the  first
approximation the  relative error of the  bias does not  depend on the
tracer population but  only on the sky density of  the sources and the
survey area.  We  therefore use as reference for  the calculations the
bias parameter  estimated for galaxies in the  range $z=0.6-1.4$ using
about 23,000  sources to $R=24$\,mag  in the AEGIS-XD  CFHTLS-D3 field
(area $\rm 1\,deg^2$).  The relative  error of the galaxy bias $\delta
b/  b$  we  estimate  for  that  sample is  about  0.14  (see  section
\ref{sec_validate}).  We further assume a surface density of X-ray AGN
after the completion of the eROSITA 4-year all-sky survey plan of $\rm
40\,deg^{-2}$.    Assuming  the  DES   $\rm  5000\,deg^2$   area  with
sufficient photometry for redshift  determination we expect a relative
AGN bias uncertainty  $\delta b/ b=0.45$.  For an  all sky survey this
number drops to about 0.16.

\subsection{The    clustering    of     X-ray    AGN    relative    to
star-forming/passive galaxies}

Currently  large  X-ray AGN  samples,  like  those  that eROSITA  will
provide,  with  sufficient  multiwavelength information  to  determine
photometric redshift  PDFs are not  available. We therefore  apply the
generalised clustering estimator to the combined AEGIS-XD and C-COSMOS
fields to  measure the clustering of  X-ray AGN and  far-IR sources in
the PEP  survey.  We  then compare these  results with  the clustering
properties of passive  galaxies selected on the UVJ  diagram.  In this
exercise  the X-rays  are used  to  identify sites  of accretion  onto
supermassive  black  holes,  the  {\it Herschel}  far-IR  data  select
star-forming galaxies  and the  UVJ diagram identifies  low (specific)
star-formation rate systems.   Therefore comparison of the environment
of  the three  samples allows  investigation of  the  relation between
black  hole growth  and the  level of  the star-formation  activity of
galaxies.

We  estimate  mean  dark   matter  halo  masses  $\rm  M_{DM}  \approx
10^{13}\,M_{\odot}\,h^{-1}$  for X-ray  AGN in  the  redshift interval
$0.6<z<1.4$ and $L_X ( \rm 2 -  10 \, keV) >10^{42} \, erg \, s^{-1}$.
This estimate is in agreement with previous determinations of the mean
dark  matter  halo mass  of  moderate  luminosity  AGN at  $z\approx1$
\citep[e.g.][]{Coil2009, Mountrichas2013}.   We also find  that within
the  error  budget  of  the  present sample,  X-ray  AGN,  IR-selected
star-forming  galaxies ($L_{IR}\ga 10^{11}\,L_{\odot}$)  and quiescent
systems live  in dark matter  halos of similar masses.   These results
show that  the clustering  of X-ray AGN  does not reveal  any relation
between accretion events onto SMBHs and star-formation.

\begin{figure}
\begin{center}

\includegraphics[height=0.9\columnwidth]{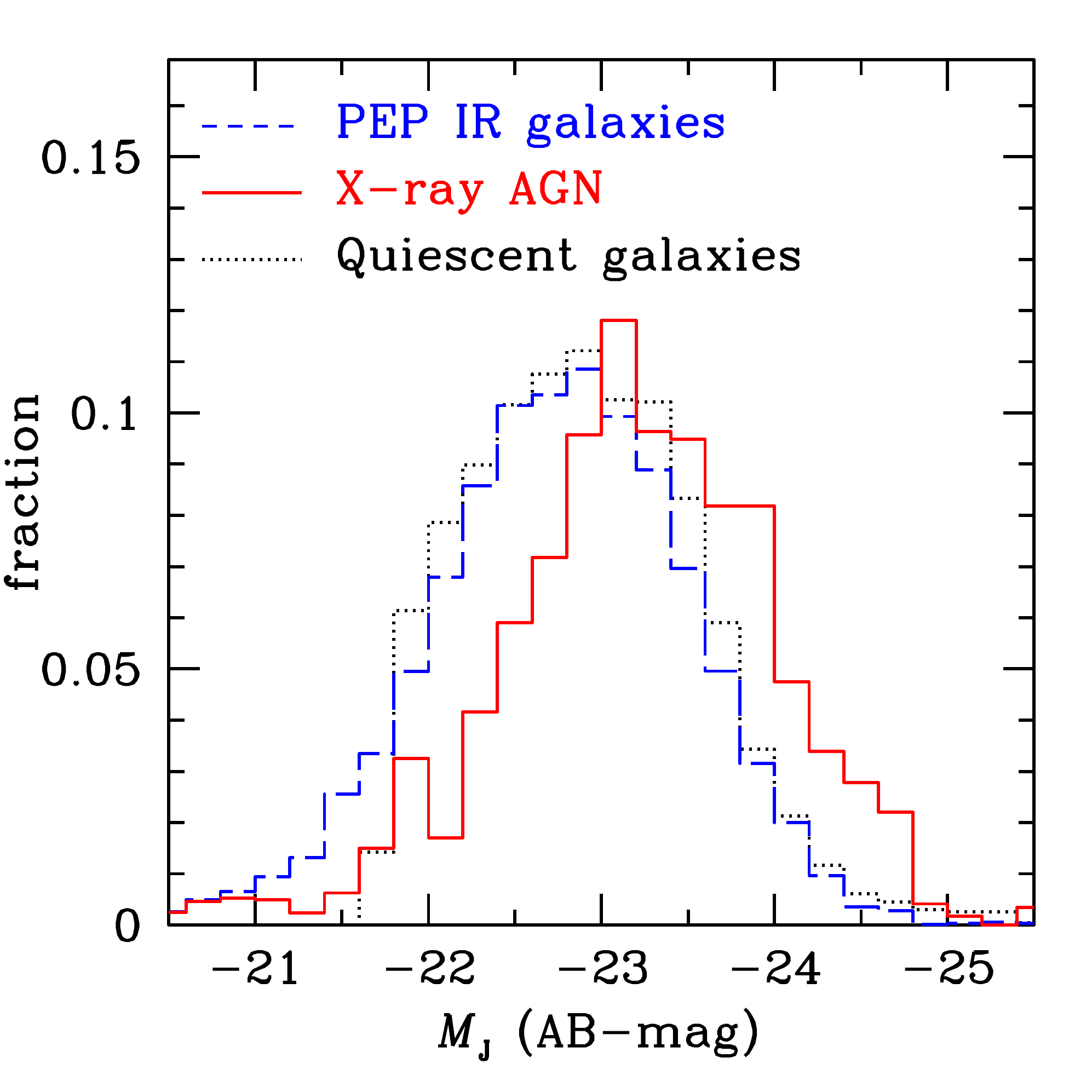}
\end{center}
\caption{$J$-band  absolute magnitude distribution  of X-ray  AGN (red
solid histogram),  far-IR selected star-forming galaxies  from the PEP
survey  (blue dashed  histogram) and  UVJ-selected  quiescent galaxies
(dotted black histogram) in the redshift interval 0.6-1.4.  Rest-frame
magnitudes  are estimated  using  {\sc k-correct}  (Blanton \&  Roweis
2007).  For sources with photometric redshifts we use the full PDFs to
determine the corresponding  $M_J$ probability distribution functions,
which  are then added  up to  produce the  distribution of  the entire
population.}\label{fig_mj_hist}
\end{figure}

\begin{figure}
\begin{center}

\includegraphics[height=0.9\columnwidth]{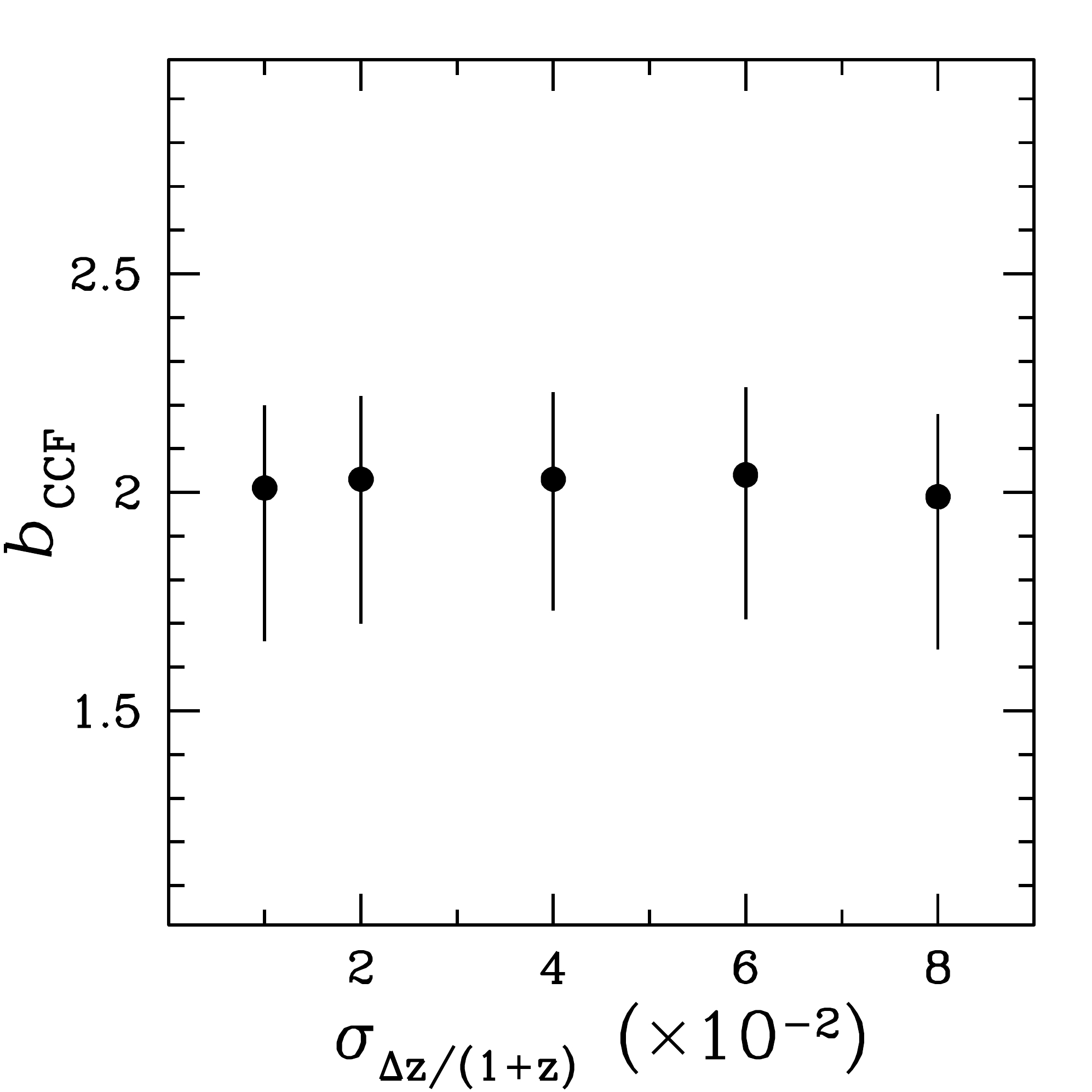}
\end{center}
\caption{AGN/galaxy cross-correlation  bias, $b_{CCF}$, as  a function
of the accuracy of the  AGN photometric redshifts, parametrised by the
dispersion $\sigma_{\Delta z / (1+z)}$. The spectroscopic redshifts of
X-ray AGN in  the C-COSMOS survey field are  convolved with a Gaussian
filter of dispersion $\sigma_{\Delta  z / (1+z)}$.  The resulting PDFs
are  then  cross-correlated  with  the photometric  redshift  PDFs  of
galaxies in the C-COSMOS survey}\label{fig_sim}
\end{figure}

This  is not  surprising given  that  all three  samples have  similar
distributions  in $J$-band  luminosity, which  is a  proxy  of stellar
mass.   Recent studies  show  that  the clustering  of  galaxies is  a
function    of   both   stellar    mass   and    star-formation   rate
\citep[e.g.][]{Li2006,     Meneux2008,     Foucaud2010,    Mostek2012,
Bielby2013}.   Star-forming galaxies  typically have  lower clustering
than  passive  ones.  At  high  stellar  masses  however, above  $\log
M_{\star}/M_{\odot}\approx  10.5$,  there  is  evidence that  the  two
populations, star-forming  and passive, have similar  dark matter halo
masses \citep{Mostek2012}.  The limit $\log M_{\star}/M_{\odot}\approx
10.5$   roughly   corresponds   to   $M_J=-21.5$\,mag   (see   section
\ref{sec_res_passive}).   Figure \ref{fig_mj_hist}  therefore suggests
that X-ray  AGN, PEP survey  far-IR sources and passive  galaxies, all
trace  massive systems  with $\log  M_{\star}/M_{\odot}\ga10.5$, which
are expected to have similar clustering properties.

A number of recent studies  speculate on the measured dark matter halo
mass of X-ray  AGN in relation to the  physical conditions under which
supermassive black  holes at the  centres of galaxies grow  their mass
\citep[e.g.][]{Allevato2011,        Mountrichas2012,       Krumpe2012,
Mountrichas2013, Fanidakis2013,  Hutsi2014}.  Our analysis  shows that
the inferred large  scale environment of X-ray AGN  is closely related
to the stellar mass of their hosts.  Galaxies with stellar mass in the
range where X-ray AGN are predominantly found, live in massive haloes,
$\log M_{DMH}/(M_{\odot}\,h^{-1})\approx13$,  independent of the level
of  star-formation  rate   (or  specific  star-formation  rate).   Put
differently,  all galaxies  that  are sufficiently  massive and  could
potentially host an  X-ray AGN, are associated with  dark matter halos
with  mass $\log  M_{DMH}/(M_{\odot}\,h^{-1})\approx13$.  We  can take
this argument  further and speculate  that claimed trends  between AGN
clustering and  luminosity or obscuration, or differences  in the mean
dark   matter  halo   mass  of   X-ray  AGN   and   UV/optically  QSOs
\citep[e.g.][]{Hickox2009,       Allevato2011,      Krumpe2013_review,
Koutoulidis2013} are  at least partially driven by  differences in the
stellar mass of the AGN hosts.

We  attempt to explore  the importance  of stellar  mass to  X-ray AGN
clustering by  splitting the AEGIS-XD and C-COSMOS  X-ray sources into
two nearly equal subsamples  at $M_J=-23.25$\,mag.  We caution that we
exclude from  these samples  X-ray sources for  which the SED  fits of
section  \ref{sec_xagn} suggest significant  AGN contamination  of the
underlying host  galaxy light.  For those  sources $M_J$ may  not be a
proxy  of  stellar  mass.   We  then  estimate  the  cross-correlation
function   with  galaxies   following  the   methodology   of  section
\ref{section:method}.  We estimate  an AGN bias of $2.1^{+0.7}_{-0.9}$
and    $1.6^{+0.5}_{-1.0}$   for    $M_J$-bright   (total    of   499;
$M_J<-23.35$\,mag; mean redshift 1.08)  and $M_J$-faint (total of 687;
$M_J>-23.35$\,mag;  mean   redshift  0.97)  X-ray   AGN  respectively.
Unfortunately,  the uncertainties  are  large and  do  not allow  firm
conclusions.  Nevertheless, taking the  estimated biases at face value
there is  tentative evidence that  the $M_J$ bright subsample  is more
clustered  than the $M_J$  faint one.   The corresponding  dark matter
haloes      for      the      two      sub-samples      are      $\log
M_{DMH}/(M_{\odot}\,h^{-1})\approx12.9$ and  12.5.  If this  result is
confirmed   by  larger   samples  it   may  offer   a  straightforward
interpretation to the clustering properties of AGN. As is the case for
galaxies, the clustering of AGN  may simply depend on the stellar mass
of their hosts.

\section{Conclusions}

We present  a novel method  for determining the  projected correlation
function of  extragalactic populations that  uses photometric redshift
PDFs  and  is  least   dependent  on  spectroscopy.   We  explore  the
performance  of the  method in  the case  of the  X-ray  AGN projected
cross-correlation function  with galaxies. We argue  that sample sizes
at least 10 times larger than spectroscopic ones are needed to recover
the clustering  signal at the  same level of accuracy.   This requires
photometric redshifts accurate at  a level better than $\sigma_{\Delta
z  /  (1+z)}\approx0.04$  for  both  AGN  and  galaxies.   Larger  AGN
photometric    redshift   errors,    e.g.     $\sigma_{\Delta   z    /
(1+z)}\approx0.08$, require a  galaxy sample with photometric redshift
dispersion better  than $\sigma_{\Delta z  / (1+z)}\approx0.01$. These
requirements place constraints  on follow-up photometric programmes of
future  large-area  X-ray  surveys,   such  as  the  eROSITA  All  Sky
Survey. Using the  projected cross-correlation function with galaxies,
we  compare the clustering  properties of  X-ray AGN,  far-IR selected
star-forming galaxies and passive systems in the AEGIS-XD and C-COSMOS
surveys.  It is found that each  of the three populations live in dark
matter       haloes        with       similar       mean       masses,
$M_{DMH}/(M_{\odot}\,h^{-1})\approx13.0$.  We argue that this is
because the  galaxies in the  three samples have similar  stellar mass
distributions, approximated by $J$-band luminosity. We hence, conclude
that the mean clustering properties of X-ray AGN are determined by the
stellar mass of their hosts.  We further speculate that claimed trends
between  AGN properties,  such  as accretion  luminosity  or level  of
obscuration, may be  driven by differences in the  stellar mass of AGN
hosts.

\section{Acknowledgments}

The  authors  are  grateful  to  the  anonymous  referee  for  helpful
comments,  M.  Krumpe  for discussions  that improved  this  paper, J.
Aird  for  providing  optical  identifications of  X-ray  sources  and
O.  Ilbert  for  making  available  photometric  redshift  Probability
Distribution Functions  for galaxies in  the COSMOS survey  field.  GM
acknowledges  financial  support  from the  Marie-Curie  Reintegration
Grant PERG03-GA-2008-230644 and the {\sc thales} project 383549, which
is jointly  funded by the European  Union and the  Greek Government in
the framework  of the  programme ``Education and  lifelong learning''.
PGP-G  acknowledges  support from  the  Spanish  Programa Nacional  de
Astronom\'ia y  Astrof\'isica under grant AYA2012-31277  This work has
made  use  of the  Rainbow  Cosmological  Surveys  Database, which  is
operated  by the  Universidad Complutense  de Madrid  (UCM), partnered
with  the  University  of   California  Observatories  at  Santa  Cruz
(UCO/Lick,UCSC).   Funding for  the DEEP2  Galaxy Redshift  Survey has
been  provided  in  part   by  NSF  grants  AST95-09298,  AST-0071048,
AST-0071198, AST-0507428,  and AST-0507483 as well as  NASA LTSA grant
NNG04GC89G.   Funding for the  DEEP3 Galaxy  Redshift Survey  has been
provided  by  NSF grants  AST-0808133,  AST-0807630, and  AST-0806732.
Based on observations obtained with MegaPrime/MegaCam, a joint project
of CFHT and CEA/DAPNIA,  at the Canada-France- Hawaii Telescope (CFHT)
which is  operated by the  National Research Council (NRC)  of Canada,
the Institut National des Sciences de l’Univers of the Centre National
de  la Recherche  Scientifique (CNRS)  of France,  and  the University
Hawaii.   This work  is based  in part  on data  products  produced at
TERAPIX  and  the  Canadian  Astronomy  Data Centre  as  part  of  the
Canada-France- Hawaii Telescope Legacy Survey, a collaborative project
of NRC and CNRS.

\bibliography{/home/age/soft9/BIBTEX/mybib}{}


\bibliographystyle{mn2e}

\appendix
\section{Testing the performance of the generalised clustering 
estimator}\label{sec_validate}

\begin{figure}
\begin{center}
\includegraphics[height=0.9\columnwidth]{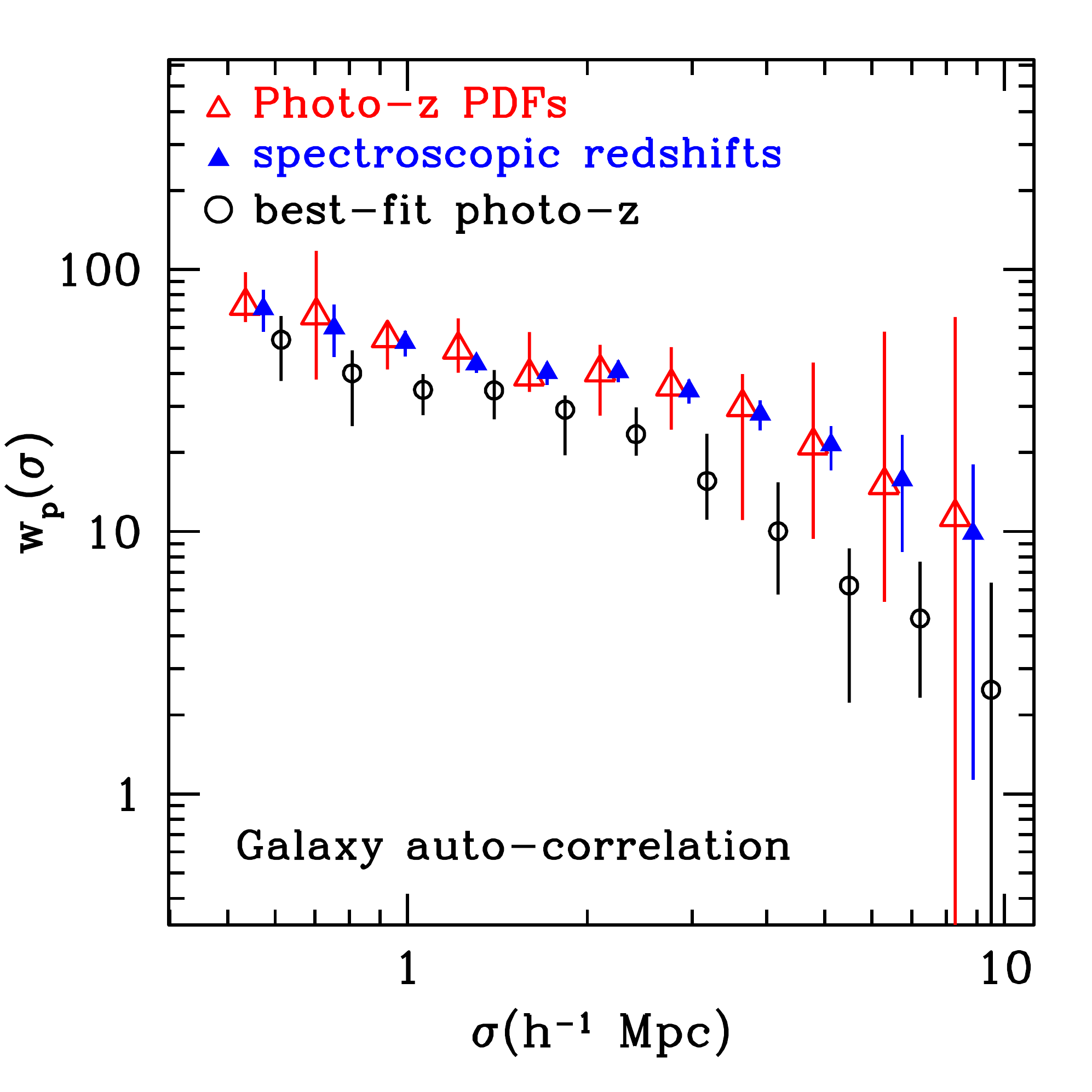}
\end{center}
\caption{Projected auto-correlation function  of galaxies plotted as a
function of scale.  The filled (blue) triangles correspond to the $\rm
w_p(\sigma)$  of spectroscopic  galaxies selected  from the  DEEP2 and
DEEP3 surveys  of the AEGIS field  with $R<24.1$ and  redshifts in the
range $0.6-1.4$.   The open (red) triangles are  estimated by applying
the   generalised   clustering    estimator   described   in   section
\ref{section:method}  to the  photometric redshift  PDFs  of CFHTLS-D3
galaxies.  The open (black) circles correspond to the auto-correlation
$\rm w_p(\sigma)$  of the same sample of  CFHTLS-D3 galaxies estimated
by replacing  the photometric redshift  PDFs with a single  value, the
photometric  redshift best-fit  solution,  and then  treating them  as
spectroscopic redshifts in the clustering analysis.  The galaxy sample
used to estimate the black open  circles and the red open triangles is
selected  to have  $R<24.1$ and  the optical  colour cuts  proposed by
Newman et al.   (2012) to exclude galaxies below  $z\approx0.6$.  As a
result, the  photometrically selected CFHTLS-D3 galaxy  sample used in
the  analysis  has  similar  properties (i.e.   redshift  and  optical
luminosity  distributions)  to  DEEP2/3 spectroscopic  galaxies.   The
errorbars  are  estimated using  $N_{JK}=8$
Jackknife regions.   For the shake  of clarity the open  red triangles
and  black circles  are offset  in  the horizontal  direction by  $\rm
\Delta\log\sigma         =          -0.02$         and         $+0.02$
respectively.}\label{fig_auto_comp}
\end{figure}

\begin{figure}
\begin{center}
\includegraphics[height=0.9\columnwidth]{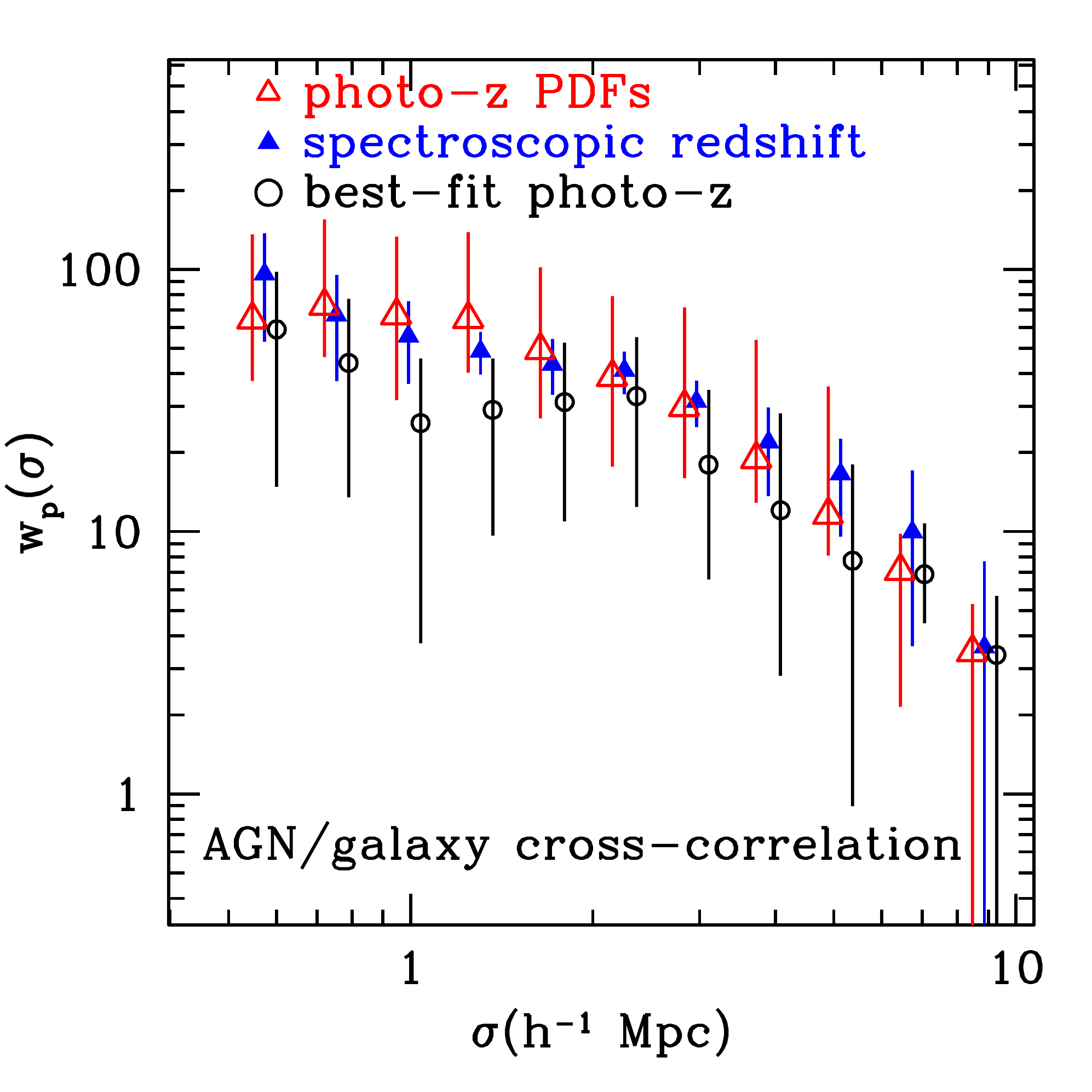}
\end{center}
\caption{ Projected  cross-correlation function between  X-ray AGN and
galaxies plotted as a function  of scale.  The filled (blue) triangles
correspond to the classic approach of estimating the cross-correlation
signal, which  requires spectroscopy for  both samples.  Spectroscopic
X-ray AGN  in the  AEGIS-XD field (total  of 148) are  correlated with
DEEP2 and  DEEP3 spectroscopic  galaxies ($\approx6,500$) in  the same
field. The  redshift range  of both samples  is limited  to $0.6-1.4$.
The open (red) triangles  are the projected cross-correlation function
estimated by  applying the generalised  clustering estimator described
in section  \ref{section:method}.  The spectroscopic  redshifts of the
148  X-ray   AGN  in  the   AEGIS-XD  field  are  replaced   by  their
corresponding  photometric redshift PDFs.   These are  then correlated
with the  photometric redshift PDFs of 23,000  CFHTLS-D3 galaxies with
$R<24.1$ and the optical colour cuts proposed by Newman et al.  (2012)
to  exclude  galaxies  below  $z\approx0.6$.    The  open  (black)
circles correspond  to the cross-correlation $\rm  w_p(\sigma)$ of the
same sample of X-ray AGN and CFHTLS-D3 galaxies estimated by replacing
the  photometric redshift PDFs  with a  single value,  the photometric
redshift best-fit  solution, and  then treating them  as spectroscopic
redshifts  in the  clustering analysis.   The errorbars  are estimated
using $N_{JK}=8$ Jackknife regions. For  the shake of clarity the open
red triangles and black circles are offset in the horizontal direction
by     $\rm     \Delta\log\sigma      =     -0.02$     and     $+0.02$
respectively.}\label{fig_cross_comp}
\end{figure}

The  generalised clustering  estimation methodology  is tested  in the
AEGIS-XD field, where extensive spectroscopy for both galaxies and AGN
is  available.  This allows  comparison with  the classic  approach of
clustering  determination   that  uses  only   spectroscopic  redshift
information.  Fig.  \ref{fig_auto_comp}, \ref{fig_cross_comp} presents
the  results  of  this  comparison  for  the  galaxy  auto-correlation
function and the AGN/galaxy cross-correlation function.

A  total of 6,500  spectroscopic DEEP2  and DEEP3  galaxies with  $R <
24.1$\,mag in  the redshift interval  $0.6-1.4$ are used  to determine
the  projected  galaxy  auto-correlation  function.  The  results  are
plotted   in  Figure   \ref{fig_auto_comp}.   Then,   the  DEEP2/DEEP3
spectroscopic sample  is replaced with 23,000  CFHTLS-D3 galaxies with
$R<24.1$\,mag and photometric redshift PDFs \citep{Coupon2009}.  These
galaxies are  selected to have similar redshifts  as the spectroscopic
subsample  by applying  the $B-R$  and  $R-I$ colour  cuts defined  by
\cite{Newman2012} to  pre-select galaxies  at $z\ga0.6$.  As  a result
the   CFHTLS-D3  photometric  redshift   galaxy  sample   has  similar
properties  (i.e.   redshift  and  luminosity  range)  to  DEEP2/DEEP3
spectroscopic  sources.    The  auto-correlation  function   of  those
galaxies is  determined using the  methodology described above  and is
plotted in  Figure \ref{fig_auto_comp}.  We estimate a  galaxy bias of
$b=1.63^{+0.07}_{-0.05}$  for  the  spectroscopic  DEEP2/DEEP3  sample
(6,500 sources) and  $b=1.72^{+0.24}_{-0.15}$ for the 23,000 CFHTLS-D3
photometric galaxies.  The  uncertainties are estimated for $N_{JK}=8$
Jackknife  regions  and  are  3--4  times  larger  for  the  CFHTLS-D3
photometric galaxy sample compared  to the spectroscopic sample.  This
is to  be expected since  the generalised clustering  methodology that
uses photometric redshifts only is geared toward large samples.  Under
the assumption that the bias  uncertainty, at a fixed galaxy magnitude
limit,  scales  as   $\sqrt{N_{gal}}$  (see  Discussion  section),  we
estimate that  a sample  with size 10  times larger than  CFHTLS-D3 is
needed to  measure the clustering  properties of galaxies at  the same
level of accuracy as the DEEP2/DEEP3 spectroscopic sample.  We caution
that this calculation  does not include the impact  of cosmic variance
in the error budget.

We further combine the  DEEP2/DEEP3 spectroscopic galaxy sample in the
range  $0.6<z<1.4$  in  the  AEGIS-XD   field  with  a  total  of  148
spectroscopic X-ray AGN in the  same redshift interval to estimate the
projected cross-correlation  function.  The result is  shown in Figure
\ref{fig_cross_comp}. The spectroscopic redshifts of the X-ray AGN are
then replaced by their corresponding photometric redshift PDFs and the
projected  cross-correlation function  with  the photometric  redshift
PDFs   of  the   23\,000  CFHTLS-D3   galaxies  is   estimated.   Fig.
\ref{fig_cross_comp} compares the  resulting signal with that obtained
using  the  spectroscopic  subsamples.   The  new  method,  that  uses
photometric redshifts  only, recovers  the clustering signal,  but, as
expected,  the  uncertainties  are  larger.  The  AGN/galaxy  bias  is
$b=1.72^{+0.12}_{-0.14}$    for   the    spectroscopic    sample   and
$b=1.84^{+0.40}_{-0.38}$   when   replacing   the   AGN   and   galaxy
spectroscopy  with  photometric  redshift PDFs  ($N_{JK}=8$  Jackknife
regions).

In  the calculations  above we  adopt  $\pi_{max} =  50$\,Mpc for  the
classic  cross-correlation and  auto-correlation  functions, based  on
spectroscopy only. In the case of the generalised clustering estimator
that uses  photometric redshift PDFs,  $\pi_{max} = 450$  and 700\,Mpc
for  the  cross-  and  auto-correlation functions,  respectively  (see
Figures \ref{fig_pimax}, \ref{fig_xi}).
 
Finally,  we  emphasize  that  the methodology  described  in  section
\ref{section:method} is necessary for an unbiased determination of the
clustering properties  of extragalactic populations  using photometric
redshifts.    Using  the   photometric   redshift  best-fit   solution
underestimates  the   clustering  signal.   Fig.   \ref{fig_auto_comp}
demonstrates this point for  the auto-correlation function of galaxies
in the AEGIS-XD field.  The  photometric redshift PDFs of galaxies are
replaced by  the the photometric  redshift best-fit solutions  and are
treated as spectroscopic redshifts. The classic approach of estimating
the correlation function  is then adopted. In this  case the resulting
signal (open circles  in Figure \ref{fig_auto_comp}) is systematically
underestimated.       Similar    conclusions   apply    to    the
cross-correlation function between X-ray  AGN and galaxies.  In Figure
\ref{fig_cross_comp}  we  also  plot the  projected  cross-correlation
function estimated by replacing  the photometric redshift PDFs of both
galaxies  and  X-ray  AGN   with  the  photometric  redshift  best-fit
solutions. These  are then treated as spectroscopic  redshifts and the
classic  approach of  estimating the  correlation function  is adopted
(black  open circles in  Figure \ref{fig_cross_comp}).   This approach
underestimates  the clustering,  although  not at  the  same level  of
amplitude as in Figure \ref{fig_auto_comp}.

\end{document}